\documentclass[lineno]{JFM-FLM_Au}

\usepackage[letterpaper, portrait, margin=1in]{geometry}
\usepackage[utf8]{inputenc}
\usepackage{amsmath} 
\usepackage{amssymb}
\usepackage{graphicx}
\usepackage{setspace}
\usepackage{parskip}
\usepackage{array}
\usepackage{booktabs}
\usepackage{hanging}
\usepackage{natbib}
\usepackage{authblk}
\usepackage{appendix}

\lefttitle{J. Birnbaum et al.}
\righttitle{Journal of Fluid Mechanics}

\title{From expansion to collapse: Bubble and continuum multiscale modeling in open-system magmas}
\author[]{Janine Birnbaum}
\author[]{Fabian B.Wadsworth}
\author[]{Anthony Lamur}
\author[]{Jackie E. Kendrick}
\author[]{Yan Lavall\'ee}
\affil[]{ Ludwig-Maximilians-Universit\"at M\"unchen}

\corresau{Janine Birnbaum, \email{J.Birnbaum@lmu.de}}

\singlespace

\begin{document}

\maketitle
\begin{abstract}
    Bubble growth in silicate melts drives significant volume expansion, which has a first order control on magma transport dynamics. When magmas are exposed to external environments, heat and volatile loss at free surfaces can reverse bubble growth, leading to shrinkage and complex feedbacks between diffusion, rheology, and flow. To resolve how magma flow controls, or is controlled by, bubble expansion, we couple a micro-mechanical model for volatile diffusion into individual bubbles, with a macro-scale thermal evolution and fluid flow of the surrounding magmatic suspension. This two-way coupling captures the co-evolution of bubble size, melt viscosity, and pressure gradients, allowing both growth and resorption to emerge naturally from local conditions. We identify distinct dynamical regimes governed by (i) bubble growth limited by (a) viscous resistance or (b) diffusion at the bubble scale, (ii) viscous transport of the suspension, (iii) outgassing through permeable porous networks and exposed magma-fluid interfaces, and (iv) thermal quenching. Across these regimes, thin, high-viscosity boundary layers arising from temperature and volatile concentration gradients play a central role in modulating flow and bubble evolution. The model is implemented in a flexible, modular numerical framework (Multiscale Vesiculation, Fluid flow, Failure, and Interaction Nonlinear model: MVFFIN) enabling extension to a wide range of systems and applications, including conduit flow and pyroclast evolution. By resolving the interplay between internal bubble dynamics and external boundary conditions, this approach provides a unified framework for understanding multiscale degassing and its impact on magmatic transport and fragmentation.
\end{abstract}

\section{Introduction} 
Volcanic eruptions, especially those with an explosive style, represent a globally-distributed geohazard \citep{Crosweller2012, Rougier2018}. Our ability to understand and potentially forecast volcanic eruptions is tied to our understanding of magma transport in the crust \citep{Degruyter2012,LaSpina2021}. At depth, magmas contain dissolved volatiles such as H$_2$O, CO$_2$, and sulfur- and halogen-based species. The most volumetrically important of these is usually H$_2$O, especially in evolved magmas that have a high potential for explosive behavior \citep{Burgisser2015, Plank2013}. As magmas rise buoyantly through the crust and to the surface, the solubility of volatile species typically decreases, resulting in the exsolution of a vapor phase into bubbles suspended within the liquid (hereafter: melt). These vapor bubbles nucleate, expand, and coalesce (i.e. vesiculate),  and may eventually form permeable, connected networks that allow gas escape out of the magma and through the conduit wall rock \citep[e.g.,][]{Eichelberger1986, Jaupart1991}. Taking these processes together, the style and intensity of volcanic eruptions is thought to be a consequence of the relative coupling efficiency of magma versus gas transport \citep{Cassidy2018}. \par 

Models for bubble-scale processes in magma should account for volatile disequilibrium \citep{Sparks1978,McIntosh2014,Watkins2017}. The full modeling of nucleation, bubble growth, resorption, and texture evolution through coalescence, Ostwald ripening, and permeable vapor flow are extremely challenging \citep{Mancini2016}. A core challenge relates to the feedbacks caused by perpetual diffusion of volatiles through the melt, creating volatile gradients which cause the melt viscosity and volatile mobilities (diffusivities) to vary locally \citep{Hess1996,Zhang2010}. The highly coupled nature of volatile diffusion, especially H$_2$O in silicate melts, necessitates numerical approaches. \par 

To make progress, we start from a shell-model for bubble growth which has been introduced and explored previously \citep[e.g.][]{Prousevitch1993,Lyakhovsky1996,Proussevitch1996,Proussevitch1998,Navon1998,Lensky2001,Coumans2020b} in which a representative bubble is initialized within a spherical shell of melt whose thickness is determined by the separation distance between neighboring bubbles, itself determined by bubble number density. We will neglect for now the processes of bubble nucleation, coalescence, and ripening processes, that have been tackled in previous works \citep[e.g.][]{Toramaru1995,Yamada2005,Toramaru2006,Castro2012a,Huber2014,Mancini2016}, but continue with the dominant process of bubble growth through water transport. Previous models have identified two regimes for controlling bubble growth: (a) a regime in which viscous forces in the melt shell resist bubble expansion, and (b) a regime in which the rate of diffusive supply of water into the bubble is the limiting process \citep{Navon1998}. The transition between these regimes depends on the bubble overpressure, radius, melt viscosity and diffusion. These fundamental insights have been essential, however, these models are limited to consideration of only the local environment in the bubble and melt shell, which then has been upscaled to all bubbles in the magma homogeneously. \par

\begin{figure}
\begin{center}
\includegraphics[width=4 in]{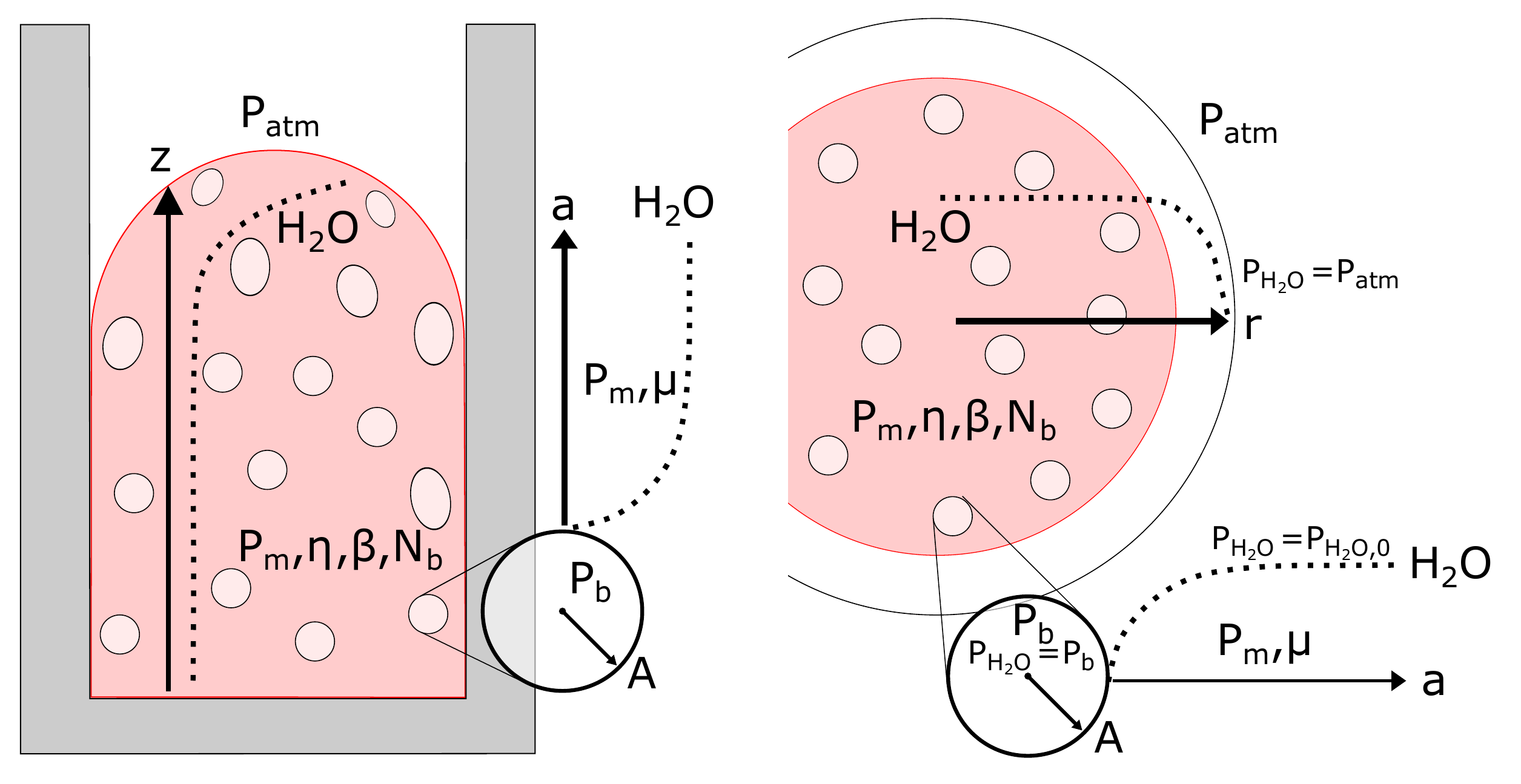}
\caption{Schematic illustrating the multiscale approach of solving water diffusion, pressures, and material properties, e.g., suspension ($\eta$) and melt ($\mu$) viscosities, and compressibility ($\beta$) at the bubble, $A$, and suspension (clast, $r$, or conduit, $z$) scales interacting with the atmosphere, for a dispersed bubble population (number density $N_b$). The water concentration in the melt surrounding the bubbles or in contact with the atmosphere is set to equilibrium with the respective partial pressure of water vapor.}
\label{fig:schematic}
\end{center}
\end{figure}

A key advance can be made if these local dynamics -- local to the bubble walls -- are coupled with larger scale dynamics, such as those that occur when a parcel of magma experiences gradients in properties on scales larger than the bubbles -– we term this ``suspension-scale dynamics''. Work on the role of suspension-scale dynamics \citep[e.g.,][]{Sparks1978,Thomas1994,Kaminski1997} demonstrates that when magmas fragment to produce pyroclasts, those pyroclasts cool through their exterior surfaces, inducing gradients in physical properties on scales larger than the bubbles. The resistance offered by the high-viscosity, cooling margins of the clasts feeds back to the local bubble-scale dynamics in the clast interiors. In this case, cooling suppresses bubble growth first in the clast margins and eventually in the clast interior once this rind becomes sufficiently thick. While the rind growth has been applied to unraveling clast thermal histories \citep[e.g.][]{Hort2000,Wright2007,Benage2014}, regimes and feedbacks between clast cooling, outgassing, and fragmentation, among other processes, on the clast interiors alluded to in earlier works have not been fully coupled and examined. 
 
In this contribution, we develop a multiscale model that explicitly accounts for the growth of bubbles at the scale of an individual bubble, based on the numerical solution of \citet{Coumans2020b}, which is then treated as representative of a bubble population in a discrete region of a suspension ($\gg 10$ bubble length scales) which can have volatile concentration or thermal gradients towards a free surface, with which the suspension exchanges mass, momentum, or energy. The numerical model developed herein represents a numerical advance in the treatment of the coupled bubble-growth and fluid-flow problem (Fig. \ref{fig:schematic}) with a modular framework called Multiscale Vesiculation, Fluid flow, Failure, and Interaction Nonlinear model (MVFFIN). The model is easily extensible to a wide range of applications, from magma flow, to other materials including synthetic glasses to bread dough and cake batter \citep{Shah1998,Vanin2009}. We provide solutions for different geometries that have both natural applications (e.g. pyroclasts, magmatic conduits), and engineered geometries relevant to controlled experiments in laboratories. In this work, we discuss several regimes that arise from the governing equations and boundary conditions, and explore in particular the controls on bubble growth due to elevated suspension viscosity due to the presence of crystals, the role of side-wall friction in conduits, and spatially variable viscosity concentrated in thin, high-viscosity boundary layers due to dehydration or cooling that resist bubble growth and lead to large overpressures. 

\begin{table}
\fontsize{10}{12}\selectfont
\centering
\begin{tabular}{|c|l|c|}
    \hline
    Variable & Description & Units \\
    \hline
    $A$, $A_0$ & Current and initial bubble radius & m \\
    $S$, $S_0$ & Current and initial melt film radius & m \\
    $a$, $a_0$ & Current and initial melt film radial coordinate & m \\
    $R$, $R_0$ & Current and initial droplet radius \textbf{or} conduit radius & m \\
    $r$, $r_0$ & Current and initial droplet radial coordinate & m \\
    $\Delta r$ & Discrete shell thickness & m \\
    $H$, $H_0$ & Current and initial magma height in conduit & m \\
    $z$, $z_0$ & Current and initial conduit vertical coordinate & m \\
    $\Delta z$ & Discrete vertical layer thickness & m \\
    $t$ & Time coordinate & s \\
    $\Delta t_n$ & Time step at iteration $n$ & s \\
    $w$ & Successive over-relaxation (SOR) relaxation factor & 1 \\

    $c$ & Volatile concentration & wt. frac. \\
    $\phi$ & Vesicularity & vol frac. \\
    $N_b$ & Bubble number density & 1/m$^{3}$ \\
    $m$ & Mass of water in bubble & kg \\
    $m_0$ & Initial mass of water in bubble & kg \\
    $m_{\text{outgas}} (t)$ & Mass of outgassed water & kg \\
    $u$ & Radial \textbf{or} vertical velocity & m/s \\
    $\dot{\varepsilon}$ & Radial \textbf{or} vertical shear strain rate & 1/s \\
    $P$ & Melt pressure & Pa \\
    $P_b$ & Bubble pressure & Pa \\
    $P_0$ & Reference pressure & Pa \\
    $\sigma_h$ & Hoop stress & Pa \\
    $T$ & Temperature & K \\

    $g$ & Gravitational acceleration & m/s$^2$ \\
    $q$ & Heat source/sink term & W/m$^3$ \\ 
    $h_c$ & Heat transfer coefficient & W/(m$^2\cdot$K) \\
    $\epsilon$ & Emissivity & 1 \\
    $\sigma$ & Stefan-Boltzmann constant & W/(m$^2\cdot$K$^4$) \\

    $D$ & Water diffusivity & m$^2$/s \\
    $\rho$ & Density & kg/m$^3$ \\
    $\rho_{\text{melt}}$ & Melt density at $P_0$ & kg/m$^3$ \\
    $\rho_{\text{H$_2$O}}$ & Water vapor density & kg/m$^3$ \\
    $\beta$ & Compressibility & 1/Pa \\
    $\beta_{\text{melt}}$ & Melt compressibility at $P_0$ & 1/Pa \\
    $G_\infty$ & Melt shear modulus at infinite frequency & Pa \\
    $Y$ & Material strength for fragmentation & Pa \\
    $\mu$ & Melt viscosity & Pa$\cdot$s \\
    $\langle \mu \rangle$ & Integrated melt viscosity & Pa$\cdot$s/m$^3$ \\
    $\mu_{\text{H$_2$O}}$ & Water vapor viscosity & Pa$\cdot$s \\
    $\eta_r$ & Relative suspension viscosity & Pa$\cdot$s \\
    $\langle \eta \rangle$ & Integrated suspension viscosity & Pa$\cdot$s/m$^3$ \\
    $\eta_\infty$ & Relative suspension viscosity at high capillarity & 1 \\
    $\eta_0$ & Relative suspension viscosity at low capillarity & 1 \\
    $\mathrm{Cc}$ & Capillary number & 1 \\
    $\Gamma$ & Surface tension & N/m \\
    $k$ & Thermal conductivity & W/(m$\cdot$K) \\
    $c_p$ & Specific heat capacity & J/(Kg$\cdot$K) \\ 
    $K$ & Gas permeability & m$^2$/s \\
    $M_{\text{H$_2$O}}$ & Molar mass of water & kg/mol \\

    \hline
\end{tabular}
\caption{List of variables and material properties with their units.}
\label{tab:variables}
\end{table}

\section{Model description}
Magmas and lavas are multiphase suspensions comprising a silicate melt phase which suspends populations of rigid crystals and vapor bubbles. The viscosity of silicate melts varies over many orders of magnitude in response to composition, including major element chemistry and volatile concentrations, and temperature, and depends weakly on pressure \citep[e.g.,][]{Hess1996,Giordano2008,DelGaudio2009}. Crystals, which typically include multiple minerals with a range of shapes and sizes, further modify the deformation behavior and rheology by resisting flow, often treated according to a relative suspension viscosity in which the viscosity is a function of melt viscosity and mineral volume fraction, as well as shape and size \citep[e.g.,]{Mader2013}. Nucleation and growth of crystals may be ongoing in magmatic systems that are below their liquidus. Similarly, vapor bubbles may be changing in shape, size, and number during magmatic processes, and contribute to the bulk rheology \citep{Pal2003,Llewellin2002b,Mader2013,Phan-Thien1997}, compressibility \citep{Rivalta2008}, and density. However, vapor bubbles differ from crystals in that the low viscosity fluid within bubbles allows for volume and shape change in response to pressure changes and shear flow in which deformation of the bulk suspension may be partially accommodated by bubbles, depending on the importance of capillary forces which resist bubble deformation \citep{Llewellin2002a}. 

When shear in the melt phase is high, either due to rapid shearing conditions, or due to concentration of shear in the presence of suspended phases \citep{Spieler2004,Vasseur2023}, deformation can exceed the ability of the silicate melt structure to viscously relax, resulting in non-Newtonian shear localization and brittle failure and fragmentation \citep{Webb1990}. Extensive fragmentation of magmas results in a transition from a melt-supported suspension, to a vapor-supported suspension of pyroclasts. This process is commonly evoked at the top of volcanic conduits and results in explosive volcanic eruptions. 

Fragmentation is promoted by rapid ascent conditions (large buoyancy) and by bubble overpressure, i.e., mass balance is such that vapor flows into bubbles or the suspension pressure drops faster than the bubble pressure can relax due to viscous flow. Such that tracking the bubble mass balance and pressure evolution is a primary objective of studying magmatic ascent. Diffusion is a primary way in which volatiles can either enter or leave bubbles through the melt, but vapor can also escape bubbles and the magma entirely (outgassing) through permeable percolation between neighboring, interconnected bubbles \citep[e.g.,][]{Mueller2005}, and out through either the top of the conduit or the surrounding wall rock \citep[e.g.,][]{Jaupart1992,Woods1994,Jaupart1998,Melnik2005b}, or through repeated fragmentation events \citep{Gonnermann2003}.

Following explosive fragmentation and ejection, pyroclasts continue to evolve in response to changing pressure and temperature conditions, which may vary rapidly in time. For example, pyroclasts that remain sufficiently hot may continue to vesiculate and fragment \citep{Wright2007,Benage2014,Namiki2021,Jones2022}. Without resolving the cooling and vesiculation history of these pyroclasts, post-fragmentation evolution may obscure the conduit conditions of interest, but with robust analysis we can unravel the full history of these samples to understand both eruptive processes and the source magma. 

In this section, we detail the numerical framework Multiscale Vesiculation, Fluid flow, Failure, and Interaction Nonlinear model (MVFFIN). First, we describe the diffusion of volatiles (water) into vapor bubbles, consistent with previous literature (Section \ref{section:bubble-scale}). We limit the scope of the model, for now, to growth by diffusion, although bubble nucleation and bubble-bubble interactions such as coalescence also contribute to the creation of vapor bubbles (vesiculation) in natural magmas. Next, we turn to the system-scale fluid flow of the magmatic suspension, including transport of both melt and vapor bubbles together under the volumetric change created by bubble growth/resorption, with material properties according to the bulk deformation behavior, and heat transfer (Section \ref{section:suspension-scale}). Suspension-scale behavior is modeled in one dimension, but in two distinct geometries: (i) spherical with radial symmetry (i.e., the only coordinate goes from the center to a free surface edge) which enables modeling of individual pyroclasts such as volcanic bombs, lapilli, or ash, and (ii) cylindrical with circumferential symmetry and no radial variability in material properties or expansion (i.e., the only coordinate goes along the axis of rotation) for modeling of conduit-type processes. Both of these geometries are additionally relevant to modeling of laboratory experiments (See e.g., \citealt{vonAulock2017,Weaver2022} for spherical-type geometries and \citealt{Colombier2026,Birnbaum2026a} for cylindrical geometry), which help improve both the interpretation of experiments and confidence of the model to capture volcanic processes for scaling to natural systems (model validation). In Section \ref{section:coupling}, we detail the numerical implementation of the multiscale coupling and temporal discretization which form the foundation of this contribution. Then, we describe two mechanisms for volatile loss (outgassing) at the suspension scale which slow or reverse bubble growth under open-system interaction with the atmosphere (Section\ref{section:open-system}). Finally, we provide criteria under which the melt or magmatic suspension would be expected to undergo brittle failure (fragmentation), which has implications for eruptive behavior and syn-eruptive modification of pyroclastic products (Section \ref{section:failure}). 

\subsection{Bubble-scale model}
\label{section:bubble-scale}
Diffusion of a volatile species $c$, in our case water, from a spherical shell of melt into a bubble is modeled via:
\begin{equation}
    \frac{\partial c}{\partial t} = \frac{1}{a^2} \frac{\partial}{\partial a} \left( a^2 D \frac{\partial c}{\partial a} \right) \: , 
\end{equation}
with a diffusivity $D$, which is dependent on composition, water content, temperature and pressure \citep{Zhang2010}, distance from the bubble center $a$, and time $t$, using the solver of \citet{Coumans2020b}. The boundary conditions at the edge of the melt film are: no flux at the far boundary, and set to equilibrium water concentration at the bubble wall according to the partial pressure of water vapor in the bubble which is simplified to be the total pressure in the bubble. \par

The pressure in the bubble, $P_b$, is a function of the mass of water, $m$, in the bubble solved by conservation of mass via the diffusion equation:
\begin{equation}
    m(t) = m_0 + 4 \pi \rho_{\text{H$_2$O}} \left( \int_{A_0}^{S_0} c(a,0) a^2 da - \int_{A_0}^{S_0} c(a,t) a^2 da \right) - m_{\text{outgas}} (t) \: , 
\end{equation}
given an initial mass of water in the bubble $m_0$, an initial water concentration $c(a,0)$, the water concentration at time $t$, $c(a,t)$, and the mass lost due to outgassing $m_{\text{outgas}} (t)$; combined with an equation of state for water vapor and the bubble volume. Given the pressure outside of the bubble, and the effect of surface tension, $\Gamma$, a bubble under- or over-pressure is calculated. The bubble volume changes in response to the pressure differential and is resisted according to the integrated melt viscosity: 
\begin{subequations}
\begin{align}
    \frac{\partial A}{\partial t} &= \frac{1}{12 A^2 \langle \mu \rangle} \left( P_b - P - \frac{2 \Gamma}{A} \right) \: , \\
    \langle \mu \rangle &= \int_{A_0}^{S_0} \frac{\mu(a) \: a^2}{(A^3 - A_0^3 + a^3)^2} da \: , 
\end{align}
\end{subequations}
for a bubble with initial radius $A_0$, current radius $A$, initial melt film radius $S_0$, melt viscosity $\mu$, and melt pressure $P$. \par 

In the original method of \citet{Coumans2020b}, the temperature and pressure path were prescribed and Matlab's ode15s ordinary differential equation solver was used to advance the solution in time according to an adaptive time step. However, for the coupled model, we need to be able to update the pressure and temperature path according to the system evolution. Accordingly, we make a few important modifications to this code to allow for running multiple times from an intermediate state: 1) rather than passing one time interval for the entire simulation, which begins from an initially flat water concentration profile, the solver is called at each time step, which begins with the initial water content and mass in the bubble defined by the spatial distribution of the previous time step, 2) the porosity calculation is now done according to the porosity of the previous time step rather than a zero porosity basis, and 3) the bubble number density evolves through time to conserve the melt volume, required by the relaxation of the assumption of zero initial porosity. The bubble growth model then uses the ode15s method to run over the duration of a single time step for the coupled model. \par

\subsection{Suspension-scale model}
\label{section:suspension-scale}
The dynamics of the bubbly suspension are treated according to the linearized compressible Navier-Stokes equations for conservation of mass and momentum in spherical:

\begin{subequations}
\begin{align}
    \frac{\partial \rho}{\partial t} &+ \frac{1}{r^2} \frac{\partial}{\partial r} \left( \rho r^2 u \right) = 0 \: , \\
    \rho \frac{\partial u}{\partial t} &= - \frac{\partial P}{\partial r}  + \frac{4}{3} \left( \frac{1}{r^2} \frac{\partial}{\partial r} \left( r^2 \eta \left( \frac{\partial u}{\partial r} - \frac{u}{2r} \right)\right) + \eta \left(\frac{1}{r} \frac{\partial u}{\partial r} - \frac{2u}{r^2} \right) \right) \: , 
\end{align}
\end{subequations}

or cylindrical confined geometry: 
\begin{subequations}
\begin{align}
    \frac{\partial \rho}{\partial t} &+ \frac{\partial}{\partial z} \left( \rho u \right) = 0 \: , \\
    \rho \frac{\partial u}{\partial t} &= - \frac{\partial P}{\partial z}  + \frac{4}{3}\left( \frac{\partial}{\partial z} \left(\eta \frac{\partial u}{\partial z} \right) - 4 \eta \frac{u}{R^2}\right) + \Delta \rho g \: ,
\end{align}
\end{subequations}
which is subject to gravity and viscous drag along the sides of the confining conduit of radius $R$, for velocity $u$, melt pressure $P$, as above, and an effective suspension viscosity $\eta$, and density $\rho$ along spatial coordinates $r$ extending from the center in the spherical case, and $z$ along the axis of symmetry in the cylindrical case. While the spherical case requires a symmetry condition on the interior boundary, we provide three options for boundary conditions in the cylindrical case: a no-flux boundary representing a constant mass of fluid (closed-system), a no-stress boundary condition which could be appropriate for modeling, for example, extrusion from a large reservoir, and an isostatic pressure gradient due to a host rock with a density contrast $\Delta \rho$. In all cases the outer boundary is treated as a free-surface. \par 

Density is provided by the constitutive relationship: 
\begin{equation}
    \rho = \left( \rho_{\text{melt}} (1-\phi) + \rho_{\text{gas}} \phi) \right) \left( 1 + \beta (P-P_0) \right) \: , 
\end{equation}
where $\rho_{\text{melt}}$ is a reference density for the melt, which is itself a function of temperature \citep{Bagdassarov1994}, $\rho_{\text{melt}}$ is the density of the water vapor in the bubble \citep{Pitzer1994}, $P_0$ is a reference pressure (taken to be atmospheric), $\phi$ is the vesicularity, and $\beta$ is the bulk magma compressibility: 
\begin{equation}
    \beta = \phi \frac{1}{P} + (1 - \phi) \beta_{\text{melt}} \: , 
\end{equation}
with a melt compressibility, $\beta_{\text{melt}}$ \citep{Malfait2011}, and assuming the bubbles behave approximately as an ideal gas. The melt compressibility is typically orders of magnitude smaller than the compressibility of the vapor phase within the bubbles. \par

The effective suspension viscosity is given by: 
\begin{subequations}
\begin{align}
    \label{eq:relative_viscosity}
    \eta  &= \mu(T,c) \eta_{r} \left( \eta_\infty +  \frac{\eta_0 - \eta_\infty}{1 + (6/5 \: \mathrm{Cc})^2} \right) \: , \\
    \eta_\infty &= (1-\phi)^{5/3} \: , \\
    \eta_0 &= (1-\phi)^{-1} \: , \\
    \mathrm{Cc} &= \left( \left| \frac{\partial u}{\partial r} \right|^2 + \left| \frac{\partial^2 u}{\partial r^2} \right|^2 \right)^{1/2} \frac{a \mu \eta_{r}}{\Gamma}
\end{align}
\end{subequations}
given the melt viscosity $\mu$, as in the bubble-scale model above, the relative viscosity contribution from suspended crystals $\eta_{r}$, which resist deformation of the suspension and can be treated using an effective medium approach, especially when smaller than the bubbles \citep[e.g.,][]{Phan-Thien1997,Mader2013,Truby2015,Birnbaum2021}, and the effective (conduit) dynamic capillary number, $\mathrm{Cc}$ \citep{Llewellin2002a,Mader2013}, where the effective strain rate is computed for the radial case, and is zero for the spherical case. Practically, the maximum effective viscosity is not permitted to exceed 10$^{12}$ Pa$\cdot$s, both for numerical stability and also because the behavior at these viscosities is essentially solid-like for deformation timescales shorter than seconds to a few minutes \citep{Dingwell1989}. \par

We solve the temperature evolution in the Lagrangian frame of reference, given in spherical coordinates: 
\begin{equation}
    \rho c_p \frac{\partial T}{\partial t} = \frac{1}{r^2} \frac{\partial}{\partial r} \left( k r^2 \frac{\partial T}{\partial r} \right) \: , 
\end{equation}
where $c_p$ is the specific heat capacity which is a function of composition, water content, temperature, and vesicularity \citep{Bagdassarov1994,Bouhifd2006,Stebbins1984}, and $k$ is the thermal conductivity of the magma, which is also a function of composition and vesicularity \citep{Bagdassarov1994}. In a cylindrical conduit: 
\begin{equation}
    \rho c_p \frac{\partial T}{\partial t} = \frac{\partial}{\partial z} \left( k \frac{\partial T}{\partial z} \right) + \frac{2}{R} q \: ,
\end{equation}
with a heat flux at the conduit margin, $q$. We simplify this term to approximate the average temperature across the conduit radius and therefore choose a simple heat flux at the margin, which in principle should depend on the radial gradient in temperature of the magma, the magma thermal properties, magma velocity, wall rock temperature and thermal properties, interaction with groundwater, and a variety of other processes that, while essentially to fully understand the thermal evolution of magma moving in conduits, is beyond the scope of this work. The boundary condition at the lava surface is imposed by either 1) a Dirichlet condition, which may be a function of time, or 2) as a flux condition with forced thermal convection across the surface and radiation. \par 

\subsection{Multiscale coupling and numerical integration}
\label{section:coupling}

We solve the suspension-scale non-dimensionalized equations (see Appendix \ref{appendix:nondimensionalization}) using finite differences on a staggered grid for velocity and pressure with a Newton-Raphson iterative scheme, and the bubble-scale model is run on the pressure nodes. The bubble-scale model is incorporated by changing the vesicularity of the suspension and the water content of the melt. Changes in vesicularity effectively constitute a volumetric source/sink term calculated through the constitutive relationship for density and the mass conservation equation. Bubble content also has a strong effect on suspension rheology. \par

The calculation of the pressure and velocity in response to bubble growth is discretized in time using the Backwards Differentiation Formula 2 (BDF2) with an adaptive time step, into which we bootstrap using the backward Euler method (BDF1). Because the bubble-scale model subdivides the time steps according to the Matlab ode15s routine, the pressure is interpolated between time steps along a linear path between the pressure at the previous time step and a forward projection for the pressure also using the method of successive over-relaxation (SOR), with an initial relaxation factor $w$ of 0.8. Properties and water contents for the micro-scale model are advanced using a Lagrangian frame of reference. Temperature is solved at a higher resolution on both the pressure and velocity nodes, also using the BDF2 scheme and thermal properties in response to bubble growth are updated in the same iterative scheme. By default, the starting two time steps are $10 \Delta t_{\text{min}}$. Then, the time step is adapted according to the previous two time steps and an acceleration or deceleration in the velocity field: 
\begin{equation}
\Delta t_{n} = \frac{1}{2}\frac{\Delta t_{n-1}^2}{\Delta t_{n-2}} + \frac{1}{2}\Delta t_{n-1} \frac{|u_{n-2}|}{|u_{n-1}|} \: , 
\end{equation}
limited so that each step cannot be smaller than $0.99 \Delta t_{n-1}$ and cannot exceed $1.01 \Delta t_{n-1}$. Additionally, we limit the time steps to maintain a monotonically increasing spatial discretization according to:
\begin{equation}
    \Delta t_{n-1} \leq -\frac{1}{2} \frac{\Delta z}{\Delta u} \: , \hspace{5mm} \text{if} \:  \frac{\Delta u}{\Delta z} < 0 \: .
\end{equation}
And finally, we do not allow the time step to be outside of a user-specified minimum and maximum unless required for stability. If the solution fails to converge after 30 iterations, it returns to the previous successfully computed time and moves forward with a time step 80\% of the previous attempt, which can result in a time step smaller than the user-defined minimum when required to find a solution,. In this case, the relaxation factor $w$ is also decreased incrementally to a minimum of 0.2. \par

\subsection{Diffusive water loss and permeable flow}
\label{section:open-system}
We allow for an open system in which volatiles are able to leave either through diffusion at the lava surface or through permeable flow of gas through the bubble network. Diffusive outgassing is handled similarly to temperature by solving the diffusion equation in a Lagrangian framework on both the pressure and velocity nodes in a spherical geometry: 
\begin{equation}
    \frac{\partial c}{\partial t} = \frac{1}{r^2} \frac{\partial}{\partial r} \left( D r^2 \frac{\partial c}{\partial r} \right) \: , 
\end{equation}
and in cylindrical (along-axis) geometry: 
\begin{equation}
    \frac{\partial c}{\partial t} = \frac{\partial}{\partial z} \left( D \frac{\partial c}{\partial z} \right) \: ,
\end{equation}
where the volatile (water) concentration at the lava surface is set to the equilibrium concentration for the activity of water dictated by partial pressure of water in the surrounding air. This water concentration at each location is used as a Dirichlet boundary condition in the bubble-scale diffusion model in place of the symmetry condition. In this implementation diffusive outgassing removes water only from the melt phase, and bubbles must resorb to reach equilibrium. \par 

Permeable flow of water vapor is governed by Darcy's law: 
\begin{equation}
   \bar{u}_{\text{H$_2$O}}  = - \frac{K}{\mu_{\text{H$_2$O}}} \frac{\partial P_b}{\partial z} \: , 
\end{equation}
where $K$ is the permeability, and $\mu_{\text{H$_2$O}}$ is the viscosity of water vapor. The model currently implements the permeability relationships of \citet{Mueller2005}, but additional models can be readily incorporated in the modular framework. We substitute to find the mass loss due to permeable flow on a per-bubble basis: 
\begin{equation}
    \frac{\partial m}{\partial t} = - \frac{3 r^2}{N_b ((r+\Delta r)^3 - r^3)} \frac{K \rho_{\text{H$_2$O}}}{\mu_{\text{H$_2$O}}} \frac{\partial P_b}{\partial r} \: , 
\end{equation}
for a spherical geometry where $m$ is the mass of water in the bubble. In cylindrical coordinates: 
\begin{equation}
    \frac{\partial m}{\partial t} = - \frac{1}{N_b \Delta z} \frac{K \rho_{\text{H$_2$O}}}{\mu_{\text{H$_2$O}}} \frac{\partial P_b}{\partial z} \: , 
\end{equation}
and in each case the pressure at the outer/top surface is set to the ambient pressure. \par 
In the simplified case of an ideal gas, the pressure and mass are directly related according to: 
\begin{equation}
    m = \frac{\frac{4}{3} \pi A^3 P_b/ M_{\text{H$_2$O}}}{\bar{R}T} \: , 
\end{equation}
where $M_{\text{H$_2$O}}$ is the molar mass of water and $\bar{R}$ is the universal gas constant. We substitute for pressure using the ideal gas law and then solve using backward Euler as an initial estimate. The pressure is then updated using the water equation of state and solved using a Newton-Raphson iteration. The gas pressure in connected bubbles is limited to be greater than or equal to the atmospheric pressure. 
\par 

For both surface degassing and permeable flow, water is removed from either the melt or bubble phases, respectively. We then update the mean water distribution for the suspension-scale model in the same time step. \par 

\subsection{Fracture and yielding}
\label{section:failure}
In some cases, the pressure of the expanding bubbles or material flow is likely to be sufficient to exceed the melt strength, resulting in localized brittle rupture. We track criteria under which rupture is expected: the melt may become sufficiently viscous and strain rates may be high enough for the material to cross through the glass transition and undergo brittle failure. High overpressure in the bubbles has been demonstrated to result in fragmentation \citep{Spieler2004b,Scheu2022} when 
\begin{equation}
    (P_b - P)\phi > Y \: , 
\end{equation}
for a material strength of $Y$. On the melt film scale, brittle failure is expected to occur when the strain rate exceeds the relaxation timescale of the melt and the stress builds to greater than the material strength \citep{Webb1990,Dingwell1996}: 
\begin{equation}
    \dot{\varepsilon} > \frac{G_\infty}{100 \mu} \: , 
\end{equation}
where $G_\infty$ is the shear modulus of the melt, with a typical value of 10$^{10}$ Pa \citep{Dingwell1989}. The tangential component of strain rate may be calculated in the melt films between bubbles, 
\begin{equation}
    \dot{\varepsilon} = \frac{u_a}{a}\: , 
\end{equation}
which at the bubble wall is: 
\begin{equation}
    \dot{\varepsilon} = \frac{1}{A}\frac{\partial A}{\partial t} \: . \\
\end{equation}
The tangential component of strain rate is higher than the radial component: 
\begin{equation}
    \dot{\varepsilon} = \frac{\partial u_a}{\partial a} \: , 
\end{equation}

in the melt flowing in the modeled direction, away from the center of a spherical droplet or along the conduit for the radial and cylindrical cases, respectively:
\begin{subequations}
\begin{align}
    \dot{\varepsilon} &= \frac{\partial u}{\partial r} \: , \\
    \dot{\varepsilon} &= \frac{\partial u}{\partial z} \: , \\
\end{align}
\end{subequations}
or perpendicular to the primary flow direction (circumferential strain) for the radial case: 
\begin{equation}
    \dot{\varepsilon} = \frac{1}{r}\frac{\partial r}{\partial t} = \frac{1}{r}u \: , \\
\end{equation}
or for friction-dominated shear flow in cylindrical conduits:
\begin{equation}
    \dot{\varepsilon} = \frac{3 u}{R} \: . \\
\end{equation}. \par

A challenge of the suspension-scale dynamics, especially in the case of diffusive gas loss or cooling, is the development of thin, high-viscosity skins or crusts on the free surface. These skins act as a barrier to continued deformation, and therefore bubble growth or absorption \citep{Kaminski1997}. These skins are commonly observed in the field \citep[e.g.,][]{Duffield1972}, in high-temperature experiments on silicate glasses \citep[e.g.,][]{vonAulock2017,Weaver2022,Schunke2026}, and in analogue experiments directed at lava flow and dome growth \citep[e.g.,][]{Fink1990,Fink1992,Griffiths1992,Griffiths1993}. Another analogue exists in the formation of crust during bread baking, on which much work has been conducted to investigate the feedbacks between bubble growth in the loaf interior (crumb) and the presence of a crust including multiscale numerical modeling \citep[e.g.,][]{deCindio1995,Zhang2006,Zhang2007,Wagner2008,Lucas2015,Nicolas2017}. Experimental studies \citep[e.g.,][]{Babin2006,Zhang2007,Wagner2008,Grenier2010} highlight that the crust acts as a mechanical barrier that resists bubble expansion, but also impedes heat and volatile loss through the surface, which can prevent permeable outgassing and insulate flow interiors. \par

If the skin or crust is dominated by elastic behavior, stress in the outer layer may be approximated according to a thin-walled hoop stress \citep[e.g.,][and references therein]{Sinclair2015}: 
\begin{equation}
    \sigma_h = \frac{r(P - P_0)}{2h} \: , 
\end{equation}
where $h$ is the finite thickness of the elastic layer. The material would be expected to rupture if the deformation conditions are such that the stress cannot be relaxes, and builds up in excess of the tensile strength of the material, producing "bread-crust" like textures \citep{Bonney1918}. If the material remains primarily fluid or plastic, the total accumulated out-of-plane strain may play an important role in plastic (necking) failure, which may be especially likely in the presence of vesicles \citep{Lister1998}. If any of these criteria are met, we may expect the viscous resistance in that layer to dramatically decrease. \par 

\section{Scaling}
\subsection{Momentum balance}
Each component of the model introduces processes with a range of possible timescales. At the scale of individual bubbles, growth regimes arise from the competition between water diffusion and melt deformation in the film surrounding the bubble, which have been extensively investigated theoretically and experimentally \citep[e.g.,][]{Proussevitch1993,Navon1998}. The corresponding timescales have frequently been reported as $\tau_{\text{diff}} \sim A^2/D$ and $\tau_{\text{melt}} \sim \mu/\Delta P_{b}$. However, the structure of the diffusion equation and numerical experimentation suggest that the characteristic length for diffusion should be the thickness of the melt film, such that at long times $\tau_{\text{diff}} \sim (S-A)^2/D$. The ratio of these timescales yields the bubble P\'eclet number $\mathrm{Pe_b}$: 
\begin{equation}
    \mathrm{Pe_b} = \frac{\tau_{\text{diff}}}{\tau_{\text{melt}}} = \frac{\Delta P_b (S-A)^2}{\mu D} \: , 
\end{equation}
where low bubble P\'eclet number indicates a regime limited by viscous resistance to expansion, and high bubble P\'eclet number is a diffusion-limited regime. In natural rhyolitic systems, both regimes are likely to occur with the high $\mathrm{Pe}$ regime promoted for low-viscosity melts (hot, wet, and less evolved) and fast decompression rates, with a mild competing effect of decompression rate on bubble number density in which $S\sim \frac{1}{2}N_b^{-1/3}$ and $N_b \sim \left( \frac{\partial P}{\partial t} \right)^{3/2}$ \citep{Toramaru2006}. As vesicularity increases, dehydration of the melt increases viscosity and the thinning melt films drive the system towards low $\mathrm{Pe}$. \par

Likewise, the Navier-Stokes equation for momentum has well-studied regimes defined by the timescales of advection, $\tau_{\text{adv}} = L/U$, viscous dissipation, $\tau_{\text{visc}} = \rho L^2/\eta$, frictional resistance along the walls, $\tau_{\text{visc}} = \rho R^2/\eta$, and gravitational acceleration, $\tau_{\text{grav}}=U/g$, and for compressible flows, the acoustic wave propagation, $\tau_{\text{sound}} = L/\sqrt{\beta \rho}$. Ratios of these timescales yield:
\begin{subequations}
    \begin{align}
        \mathrm{Re} &= \frac{\tau_{\text{visc}}}{\tau_{\text{adv}}} = \frac{\rho L U}{\eta} \: , \\
        \mathrm{Fr}^2 &= \frac{\tau_{\text{grav}}}{\tau_{\text{adv}}} = \frac{U^2}{g L} \: , \\ 
        \mathrm{Ma} &= \frac{\tau_{\text{sound}}}{\tau_{\text{adv}}} = \frac{U}{\sqrt{\beta \rho}} \: , 
    \end{align}
\end{subequations}
where the Reynolds number $\mathrm{Re}$ compares inertial and viscous forces which we assume to be low (viscosity-dominated), which is a valid assumption for most volcanological applications, with the exception of some fast-moving lava flows \citep[e.g.,][]{Dietterich2022} and lava fountains and plumes \citep{Sparks1983, LaSpina2021, Griffiths2000}. The Froude number $\mathrm{Fr}$ compares inertia with gravity, and is applied only in the cylindrical conduit case. From compressiblity we get the Mach number $\mathrm{Ma}$ which compares flow velocity to the speed of sound in the material, which is small for coherent flows, but may be significant in volcanological problems near a decompression-driven fragmentation front which excavates the conduit. \par 

In the cylindrical confined conduit case, we also get the aspect ratio of length to radius, $\mathrm{R_A} = L/R$ and find the Poiseuille number: 
\begin{equation}
    \mathrm{Ps} = \frac{\mathrm{Fr}^2 \mathrm{R_A}^2}{\mathrm{Re}} \: , 
\end{equation} 
which compares the frictional resistance provided by the conduit walls to buoyancy, that is the characteristic length scale for the Reynolds number becomes the conduit diameter rather than flow length. Similarly, if Poiseuille flow is driven by an applied pressure gradient rather than buoyancy, the relevant scaling is $\mathrm{R_A}^2$. In this work, we choose to neglect the buoyancy-driven segregation of bubbles from the melt, which depends on the Stokes number $\mathrm{St}$: 
\begin{equation}
    \mathrm{St} = \frac{\tau_{adv}}{\tau_{drag}} = \frac{\pi a^2 \rho g}{3 \eta U} \: , 
\end{equation}
such that large $\mathrm{St}$ suggests fast rise of bubbles compared to the advection timescale $\tau_\text{drag}= \pi a^2 \rho g/(3 \eta)$. While this process may be important for certain low viscosity lava flows, Stokes numbers are typically low during conduit ascent and for the fragmental systems of evolved magmas that we target here. \par  

\begin{figure}
\begin{center}
\includegraphics[width=4 in]{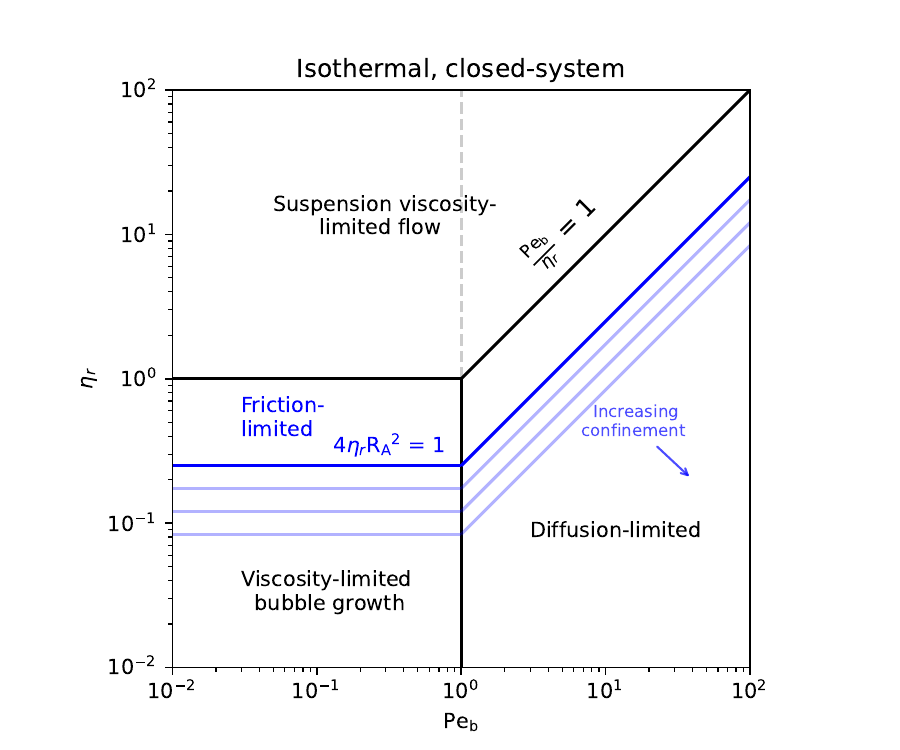}\\
\caption{Regime diagram for iso-thermal, closed system flows contrasting different bubble growth regimes and the importance of the resistance provided by viscous deformation of the suspension both with and without friction in confined conduits.}
\label{fig:isothermal_closed_confined_regime}
\end{center}
\end{figure}

We can compare the dominant rates across the bubble- and suspension spatial scales to find regimes of behavior for the coupled model:
\begin{subequations}
\begin{align}
        \frac{1}{\eta_r} &= \frac{\tau_{\text{melt}}}{\tau_{\text{adv}}} = \frac{U \mu}{L \Delta P_b}  = \frac{\mu}{\eta} \: ,  \: &\text{for} \: \mathrm{Pe_b} << 1 \: ,  \\
         \frac{1}{\delta^2}\mathrm{Sc} &= \frac{\tau_{\text{diff}}}{\tau_{\text{visc}}} = \frac{(S-A)^2 \eta}{L^2 D \rho} \: ,  \: &\text{for} \: \mathrm{Pe_b} >> 1 \: . 
\end{align}
\end{subequations}
As in the suspension-scale momentum conservation above, the choice of suspension-scale pressure dominated by viscous effects gives $\Delta P \sim \eta U/L$, and we assume that the pressure driving growth is of the same order as the bubble overpressure. This reduces the first relationship to the ratio of the melt viscosity to the suspension viscosity, also called the relative viscosity, $\eta_r$ 
 (Fig. \ref{fig:isothermal_closed_confined_regime}). In the second, we take the ratio of the diffusion and viscous timescales which gives a modified Schmidt number, $\mathrm{Sc}$, which includes the square of ratio of the melt film thickness and the system length scale. Naturally, we can compare opposite pairs which can then be expressed as ratios of the same numbers. In the friction-dominated case, the pressure term should follow $\Delta P \sim \eta U L/R^2$, which yields the relationships as above with an additional $\mathrm{R_A}^2$ term for the melt viscosity-dominated bubble growth. \par

\subsection{Vapor mass balance}
Diffusive outgassing at the free surface also produces a timescale of diffusion $\tau_{\text{outgas}} = L^2/D$ with now the characteristic length scale defined by the droplet radius or column height in a conduit instead of the bubble-scale. To find the timescale of permeable flow, we choose a characteristic timescale in which $\bar{u}_{\text{H$_2$O}} \sim L/\tau_{\text{perm}}$ resulting in $\tau_{\text{perm}} = L^2 \mu_{\text{H$_2$O}}/K \Delta P_b$. We compare the bubble growth timescales to gas loss due to diffusive outgassing: 
\begin{subequations}
\begin{align}
        \mathrm{\delta}^2 &= \frac{\tau_{\text{outgas}}}{\tau_{\text{diff}}} = \frac{L^2}{(S-A)^2} \: ,  \: &\text{for} \: \mathrm{Pe_b} >> 1 \: , \\
        \mathrm{Pe_s} &= \frac{\tau_{\text{outgas}}}{\tau_{\text{melt}}} = \frac{\Delta P_b L^2}{\mu D} \: ,  \: &\text{for} \: \mathrm{Pe_b} << 1 \: ,\end{align}
\end{subequations}
\begin{subequations}
and gas loss due to permeable flow:
\begin{align}
        \frac{1}{\mathrm{Sh}} &= \frac{\tau_{\text{perm}}}{\tau_{\text{diff}}} = \frac{L^2 \mu_{\text{H$_2$O}} D}{(S-A)^2 K \Delta P_b} \: ,  \: &\text{for} \: \mathrm{Pe_b} >> 1 \: , \\
        \frac{\mathrm{\lambda_{\text{H$_2$O}}}}{\mathrm{Da}} &= \frac{\tau_{\text{perm}}}{\tau_{\text{melt}}} = \frac{L^2}{K} \frac{\mu_{\text{H$_2$O}}}{\mu}\: ,  \: &\text{for} \:  \mathrm{Pe_b} << 1 \: .
\end{align}
\end{subequations}
\par 

\begin{figure}
\begin{center}
\includegraphics[width=\textwidth]{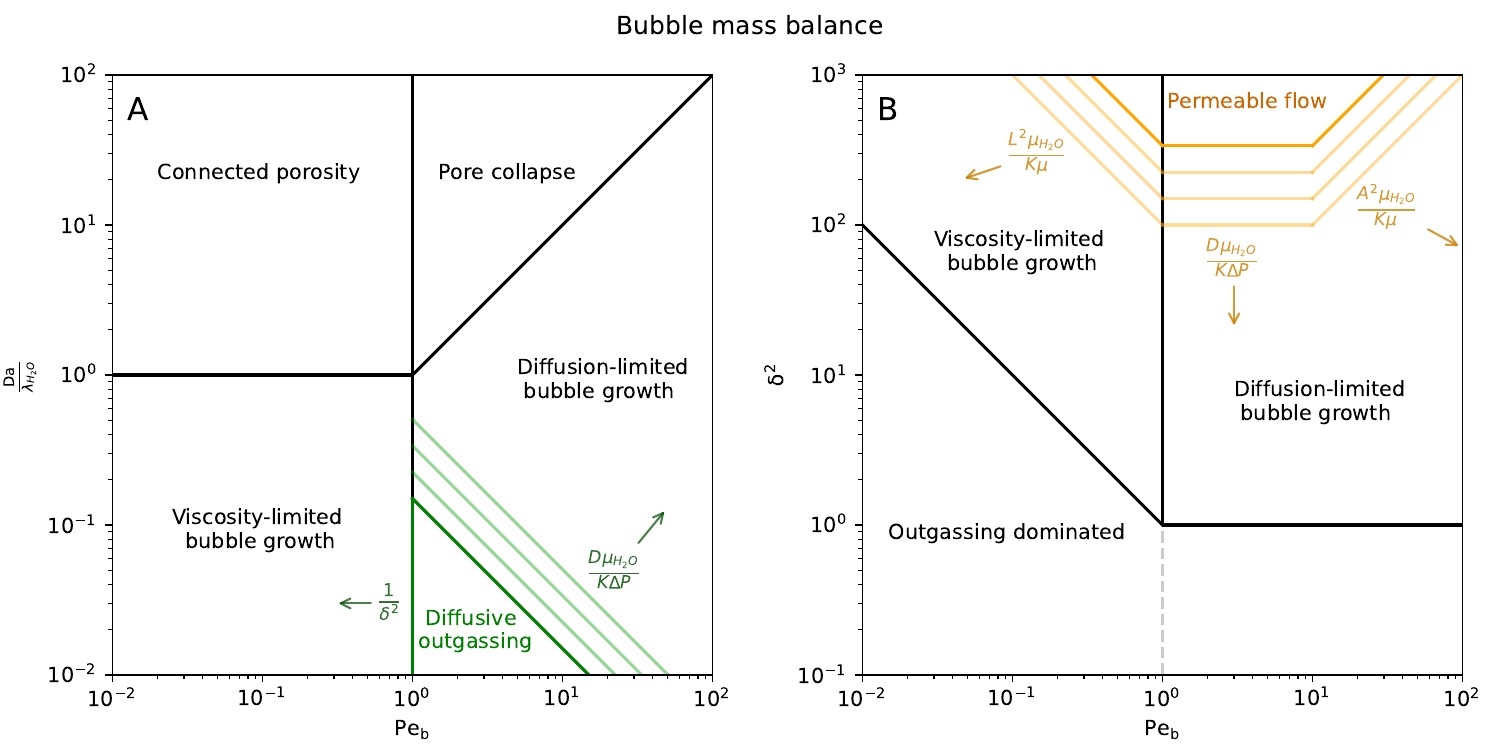}\\
\caption{Regime diagram for iso-thermal, open system flows contrasting different bubble growth regimes and the influence of either A) diffusive outgassing at the surface or B) permeable flow.}
\label{fig:bubble_mass_balance_regime}
\end{center}
\end{figure}

We find another P\'eclet number, this time for the suspension-scale system which is related to the bubble P\'eclet number by $\mathrm{Pe_s} = \mathrm{\delta}^2\mathrm{Pe}$, where $\mathrm{\delta}$ is the ratio of the suspension length-scale to the melt film thickness (Fig. \ref{fig:bubble_mass_balance_regime}). Implicit in the multiscale model is the assumption that $\mathrm{\delta}>>1$ such that outgassing via surface diffusion for the clast interior is slow with respect to the bubble response time, and can be significant only in thin rinds. Given typical water diffusivities in magmas, this scaling highlights that diffusive outgassing is relevant only for small systems (e.g., ash to lapilli sized clasts) unless the timescale of interest is very long (at least years to centuries for length scales of a meter). For magma ascent in conduits, diffusive water transport is likely not important, in either the along-conduit direction as we model here, or indeed in the radial direction ($\tau_{\text{outgas}}=R^2/D$). However, diffusive outgassing through an upper free surface may be important for experimental systems, typically order millimeters to centimeters, which makes this model well-posed to simulate and evaluate differences between experimental and natural scales. \par 

The ratio of timescales of permeable flow to bubble deformation is the ratio of water vapor to silicate melt viscosity, $\mathrm{\lambda_{\text{H$_2$O}}} = \mu_{\text{H$_2$O}}/\mu$ and the inverse of the Darcy number, $\mathrm{Da}=K/L^2$. The relative contributions of water vapor mass transfer from permeable flow and diffusion into the bubbles defines a Sherwood number, which can also be expressed as a combination of the preceding numbers: $\mathrm{Sh} = \mathrm{\lambda_{\text{H$_2$O}}}/(\mathrm{Da}\mathrm{Pe_b})$. \par 

\subsection{Heat balance}
\begin{figure}
\begin{center}
\includegraphics[width=4 in]{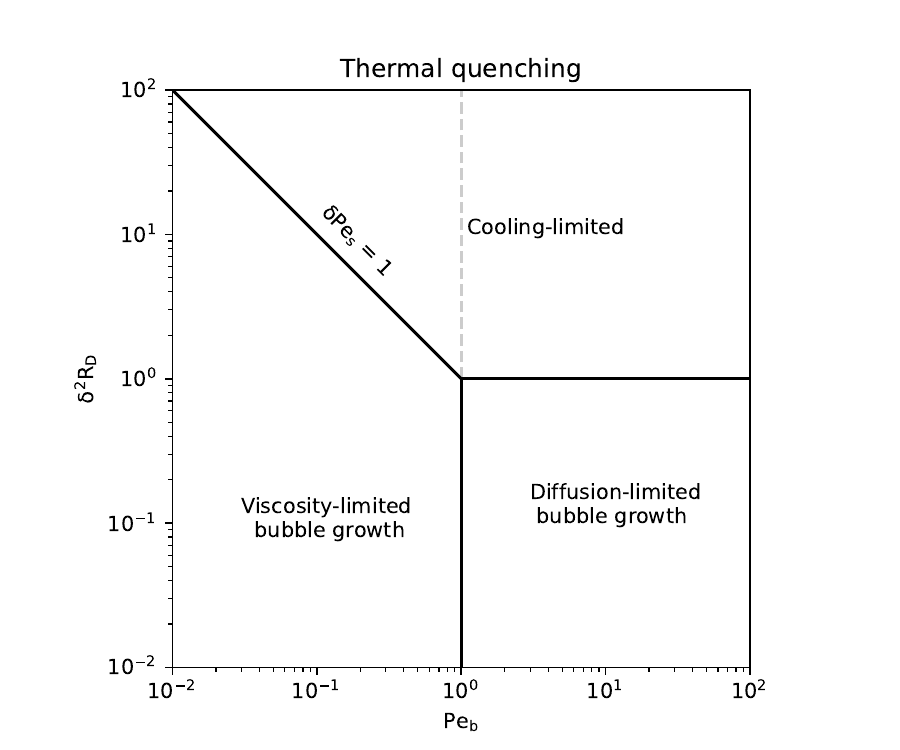}\\
\caption{Regime diagram for cooling.}
\label{fig:thermal_quenching_regime}
\end{center}
\end{figure}

Conservation of heat in the Lagrangian framework produces only a diffusion timescale for thermal conduction, $\tau_{\text{cond}} = L^2\rho c_p/k$, and we neglect thermal gradients at the bubble-scale. We consider the case of the competition between diffusive outgassing and thermal conduction given the important role temperature plays on the diffusion of water in the melt: 
\begin{equation}
    \mathrm{R_D} = \frac{\tau_{\text{cond}}}{\tau_{\text{diff}}} = \frac{\rho c_p D}{k} \: , 
\end{equation}
which yields the ratio of the water- and thermal diffusivities $\mathrm{R_D}$ (Fig. \ref{fig:thermal_quenching_regime}). In comparison with the timescale of flow we arrive at another P\'eclet number:
\begin{equation}
    \mathrm{Pe_T} = \frac{\tau_{\text{cond}}}{\tau_{\text{adv}}} = \frac{L U \rho c_p}{k} \: ,
\end{equation}
and, we consider the competition between thermal diffusion and bubble growth: 
\begin{subequations}
\begin{align}
        \mathrm{\delta}^2\mathrm{R_D} &= \frac{\tau_{\text{cond}}}{\tau_{\text{diff}}} = \frac{L^2 \rho c_p D}{(S-A)^2 k} \: ,  \: &\text{for} \: \mathrm{Pe_b} >> 1 \: , \\
        \mathrm{\delta}\mathrm{Pe_T} &= \frac{\tau_{\text{cond}}}{\tau_{\text{melt}}} = \frac{L^2 \rho c_p \Delta P_b}{k \mu} \: ,  \: &\text{for} \: \mathrm{Pe_b} << 1 \: . \\
\end{align}
\end{subequations}
We also include the possibly boundary conditions for heat transfer at the surface compared to conduction in the interior to find the Biot number: 
\begin{equation}
    \mathrm{Bi} = \frac{h_c L}{k} \: , 
\end{equation}
\par
for a heat transfer coefficient, $h_c$, and the ratio of heat lost due to radiation and conduction: 
\begin{equation}
    \mathrm{Sk} = \frac{\mathrm{Pe_T}}{\mathrm{Bo}} = \frac{\epsilon \sigma T^3 L}{k} \: ,
\end{equation}
where $\mathrm{Sk}$ is the Stark number and $\mathrm{Bo}$ is the Boltzmann number, for an emissivity $\epsilon$, the Stefan-Boltzmann constant $\sigma$, and the characteristic temperature $T$.

\begin{table}[h!]
\fontsize{10}{12}\selectfont
\centering
\renewcommand{\arraystretch}{1.8} 
\begin{tabular}{|c|l|c|}
    \hline
    Abbreviation & Dimensionless Number & Time- or lengthscales \\
    \hline
    $\mathrm{Bi}$ & Biot number & $\frac{\tau_\text{forced adv}}{\tau_\text{cond}}$ \\
    $\mathrm{Bo}$ & Boltzmann number & $\frac{\tau_\text{radiation}}{\tau_\text{adv}}$ \\
    $\mathrm{Cc}$ & Capillary number & $\frac{\text{Shear forces on bubbles}}{\text{Surface tension forces}}$ \\
    $\mathrm{Da}$ & Darcy number & $\frac{\tau_\text{melt}}{\tau_\text{perm}} \frac{1}{\lambda_{\text{H$_2$O}}}$\\
    $\mathrm{Fr}$ & Froude number & $\left(\frac{\tau_\text{grav}}{\tau_\text{adv}}\right)^{1/2}$ \\
    $\mathrm{Ma}$ & Mach number & $\frac{\tau_\text{sound}}{\tau_\text{adv}}$ \\
    $\mathrm{Pe_b}$ & Bubble P\'eclet number & $\frac{\tau_\text{diff}}{\tau_\text{melt}}$ \\
    $\mathrm{Pe_s}$ & Suspension P\'eclet number & $\frac{\tau_\text{outgas}}
    {\tau_\text{melt}}$ \\
    $\mathrm{Pe_T}$ & Thermal P\'eclet number & $\frac{\tau_\text{cond}}{\tau_\text{adv}}$ \\
    $\mathrm{Ps}$ & Poiseuille number & $\frac{\tau_\text{grav}}{\tau_\text{visc}} {R_A}^2$\\
    $\mathrm{Re}$ & Reynolds number & $\frac{\tau_\text{visc}}{\tau_\text{adv}}$ \\
    $\mathrm{R_A}$ & Conduit aspect ratio & $\frac{L}{R}$ \\
    $\mathrm{R_D}$ & Water to thermal diffusivity ratio & $\frac{\rho c_p D}{k}$ \\
    $\mathrm{Sc}$ & Schmidt number & $\frac{\tau_\text{diff}}{\tau_\text{visc}} \delta^2$ \\
    $\mathrm{Sh}$ & Sherwood number & $\frac{\tau_\text{diff}}{\tau_\text{perm}}$ \\
    $\mathrm{Sk}$ & Stark number & $\frac{\tau_\text{cond}}{\tau_\text{radiation}}$ \\
    $\mathrm{St}$ & Stokes number & $\frac{\tau_\text{adv}}{\tau_\text{drag}}$ \\
    $\mathrm{\delta}$ & Suspension to melt film length ratio & $\frac{L}{S-A}$ \\
    $\mathrm{\eta_r}$ & Relative suspension viscosity & $\frac{\eta}{\mu}$ \\
    $\mathrm{\lambda_{\text{H$_2$O}}}$ & Water vapor to melt viscosity ratio & $\frac{\mu_{\text{H$_2$O}}}{\mu}$ \\

    \hline
\end{tabular}
\caption{List of dimensionless numbers and ratios.}
\label{tab:numbers}
\end{table}

\section{Results}
\subsection{Comparison to previous bubble-scale models}
\begin{figure}
\begin{center}
\includegraphics[width=4 in]{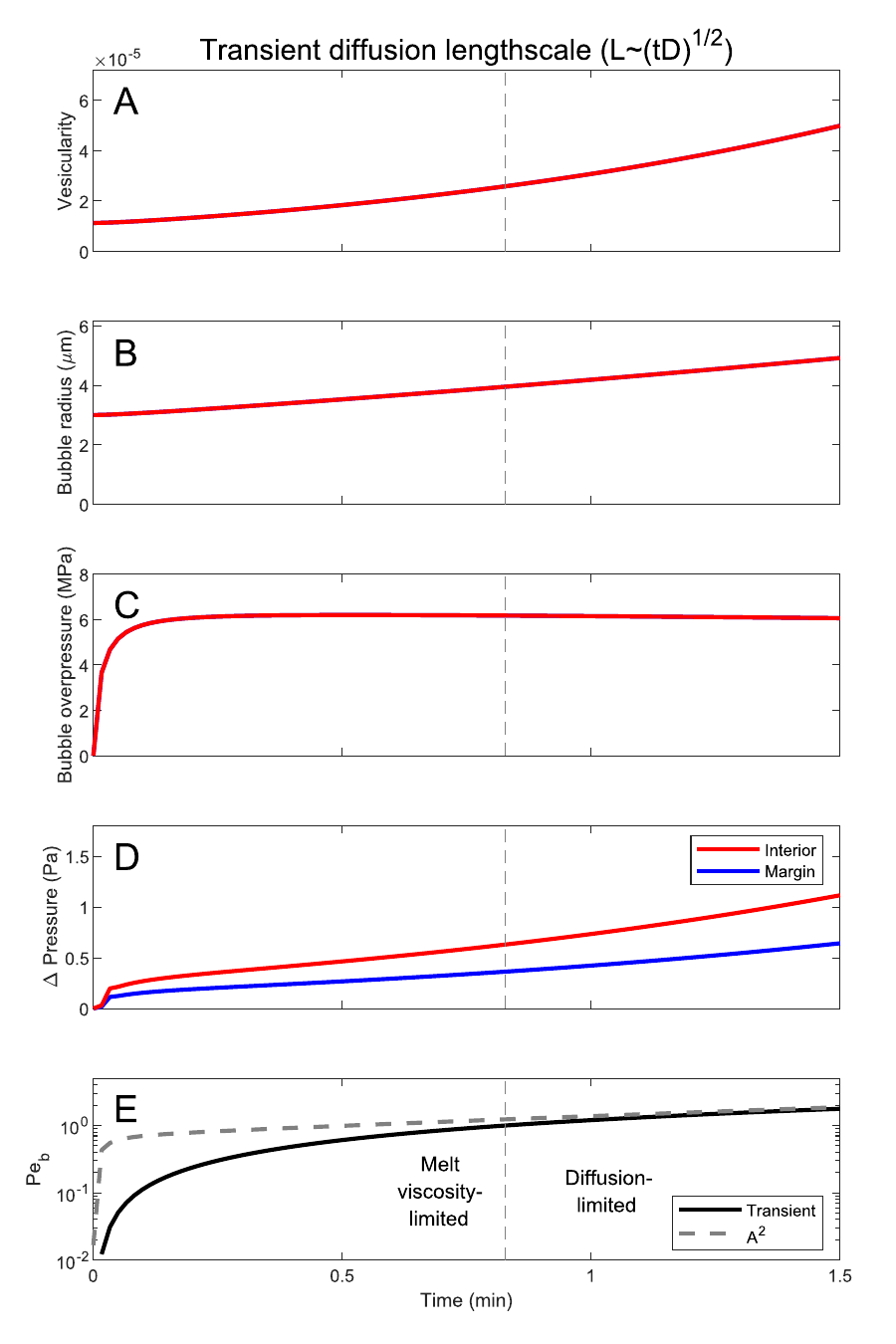}\\
\caption{Simulation results for the isothermal, closed-system regime across the transition from transient diffusion-limited to viscosity-limited regimes showing A) vesicularity, B) bubble radius, C) bubble overpressure, D) pressure above atmospheric through time, in both the margin (blue) and interior (red) of a spherical droplet. E) $\mathrm{Pe_b}$ calculated using a transient length scale ($\sim\sqrt{tD}$, black) and the bubble radius ($A$, gray, dashed). The dashed vertical line in all panels indicates the time at which $\mathrm{Pe_{b,transient}}$=1. $\mathrm{Pe_{b}}$ calculated using the melt film thickness is much greater than 1.}
\label{fig:small_bubbles}
\end{center}
\end{figure}

We first consider the simplified scenario of an isothermal ($\mathrm{Pe_T} \rightarrow \inf$), closed system ($\mathrm{Sh} \rightarrow 0$). To illustrate the dynamics modeled here, we choose a canonical magma system of a rhyolite matching the composition from Guagua Pichincha volcano from \citet{Wright2007} that begins with an initial water content (1 wt\%) which is higher than the solubility at the chosen temperature (720 $^\circ$C) and ambient pressure (atmospheric) using the solubility model of \citet{Liu2005}, the water diffusivity model for metaluminous systems of \citet{Zhang2010}, and viscosity model of \citet{Hess1996}. The model is initialized with 10$^{11}$ 1/m$^3$ bubbles of 3 $\mu$ radius and no overpressure in the bubbles, i.e. the initial pressure is the Laplace pressure given a surface tension of 0.22 N/m \citep{Bagdassarov2000}. Initially, to isolate the bubble-scale effects, we begin with a spherical magma body of 5 cm radius, and impose an artificial relative viscosity of 0.1. While we make these choices in order to close the model and explore the dynamics it predicts, we stress that the model itself is general and not specific for a given magma/melt type/composition; our downloadable code supplement is specifically designed to be modular such that these specificities can be defined by a user and tailored to a use case. Under these conditions, we can reproduce the well-known result in which an initially small bubble rapidly develops an over-pressure, driving bubble expansion (Fig. \ref{fig:small_bubbles}). \par

In this early stage of bubble growth, the entire melt thickness does not participate in the water diffusion, but instead the critical length for diffusion is transient and scales with $\sqrt{Dt}$ \citep{Proussevitch1993,Chernov2014}. There is no marked change in behavior at $\mathrm{Pe_{b,transient}}$=1, although there is a transition from increasing to decreasing bubble overpressure at a critical value of $\sim10^{-1}$. In this particular case, $\mathrm{Pe_{b,transient}}$=1 happens to coincide with the region in which the classic definition of the bubble P\'eclet number, $\mathrm{Pe_{b,A^2}} = \tau_{\text{diff}}/\tau_{\text{melt}} = \Delta P_b A^2/(\mu D)$ is also 1, but this is not universally the case and depends on the initial bubble pressure and water content in the melt. The similarity in these two values at this critical range may have contributed to confusion in the literature about the appropriate characteristic length scale; instead, we would argue that this transient length scale approach gives a better measure of the earliest bubble behavior. At longer times, $\sqrt{Dt}$ must eventually approach the melt film thickness, where this transition is primarily sensitive to the bubble number density rather than the bubble size. The transient diffusion-dominated regime in which the initially sharp diffusivity profile only applies for short durations (under these conditions less than one minute), accounts for an extremely small amount of vesiculation (less than a doubling of the initial bubble radius), and is only relevant for near-instantaneous changes in water solubility.  \par

\begin{figure}
\begin{center}
\includegraphics[width=4in]{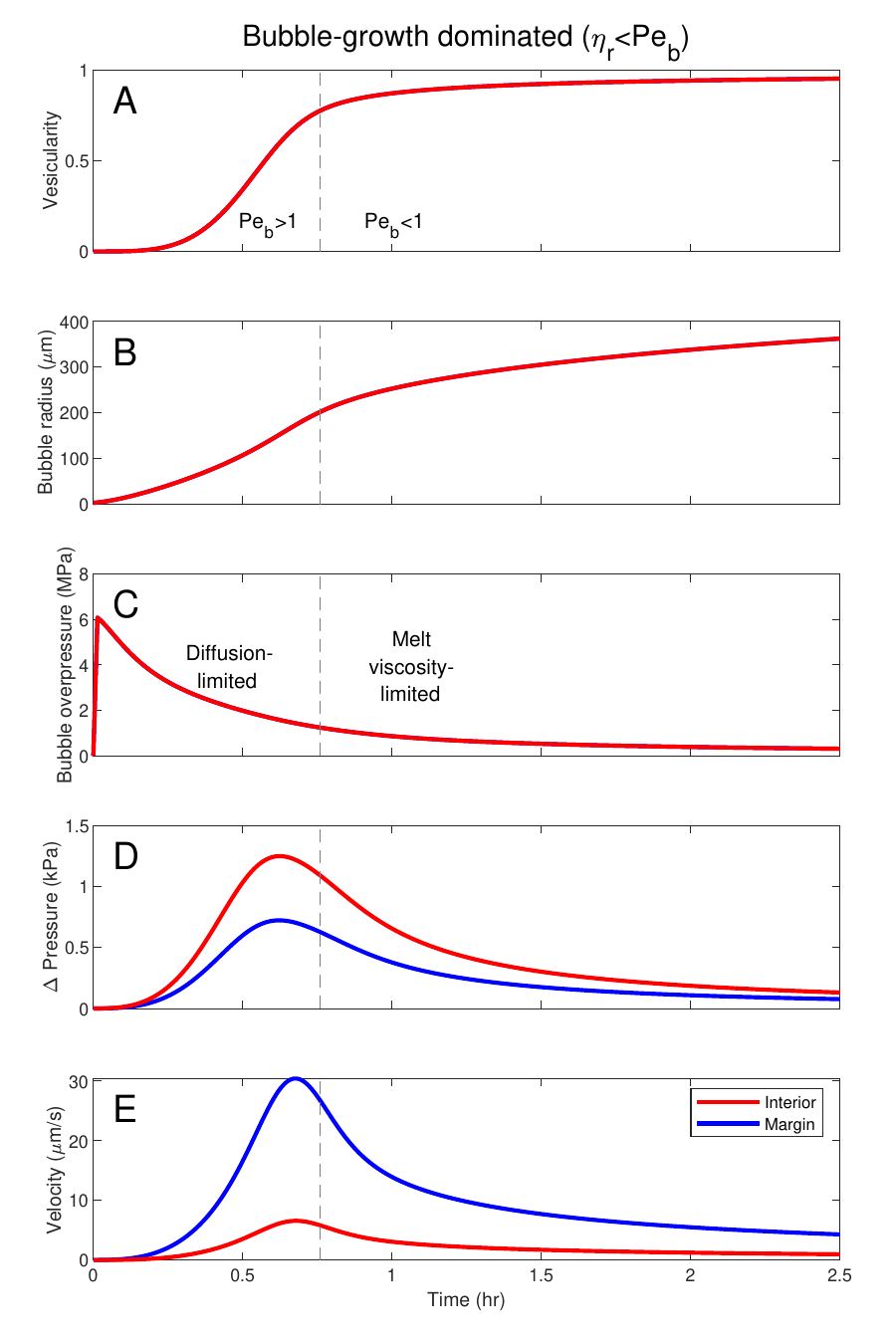}\\
\caption{Simulation results for the isothermal, closed-system, diffusion-limited regime showing A) vesicularity, B) bubble radius, C) bubble overpressure, D) pressure above atmospheric, and E) velocity through time (left) in both the margin (blue) and interior (red) of a spherical droplet, and radial distance (right) for a sphere initially 5 cm in radius. The dashed vertical line in all panels indicates the times at which $\mathrm{Pe_b}=1$}
\label{fig:bubble_growth_dominated}
\end{center}
\end{figure}

\begin{figure}
\begin{center}
\includegraphics[width=4 in]{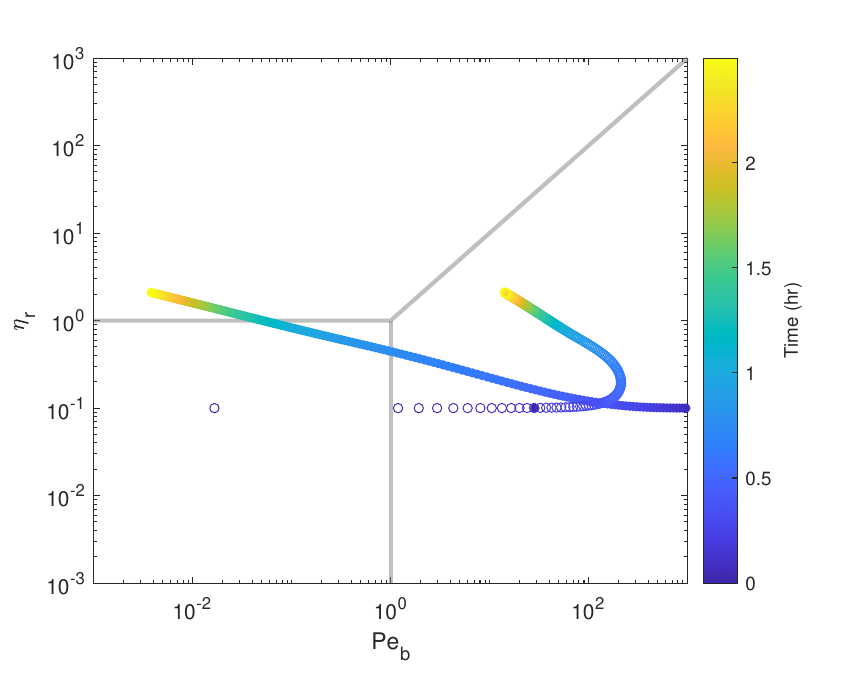}\\
\caption{Evolution of the $\mathrm{Pe_b}$ and $\eta_r$ for the simulation in Fig. \ref{fig:bubble_growth_dominated}. Color indicates the simulation time. Symbols contrast the calculation of $\mathrm{Pe_b}$ for a characteristic length scale of $S-A$ (closed circles) and $A$ (open circles). The simulation begins with an imposed relative velocity of 0.1 to ensure the suspension offers no additional resistance, but the vesiculation eventually drives the simulation to $\eta_r>1$.}
\label{fig:bubble_growth_regime}
\end{center}
\end{figure}

If we extend this simulation in time (Fig. \ref{fig:bubble_growth_dominated}), we instead calculate the bubble P\'eclet number using the melt film thickness. The diffusion-limited regime persists through much of the vesiculation history. As the melt slowly dehydrates and thins, the melt viscosity increases and the length scales shrink such that $\mathrm{Pe_b}$ becomes less than 1. For this definition, we find that $1<\mathrm{Pe_b}<5$ coincides with a change in system behavior including a reduction in the bubble radius growth rate and a corresponding transition from increasing to decreasing velocity, confirming the choice of length scales. At this same time, the classical definition of $\mathrm{Pe_{b,A^2}}$ is decreasing from a maximum of $\sim10^2$ (Fig. \ref{fig:bubble_growth_regime}), but does not indicate a regime change.  \par

\subsection{Relative viscosity}
The advance of this work lies in resolving the competition between micro-scale bubble growth and macroscopic flow. From the scaling above, we see that the timescale of viscosity-dominated bubble growth and the timescale for macroscopic flow are directly related by only the relative suspension viscosity. That is, when the suspension viscosity is less than or equal to the melt viscosity, the macroscopic flow offers no significant additional resistance to bubble expansion and only the micro-scale model is important (Fig. \ref{fig:isothermal_closed_confined_regime}). In practice, the pressure gradients driving flow and bubble growth are not precisely the same and the suspension must contribute a small resistance depending on the geometry, leading to a modest increase in total pressure in the interior of the magma parcel, which becomes more important with increasing suspension viscosity. \par 

The regime in which the suspension viscosity is significantly lower than the melt viscosity occurs in some natural systems, with the most important case being at high vesicularity and high capillary regime (Eq. \ref{eq:relative_viscosity}). We simulate the evolution of such a flow in an diffusion-limited regime (Fig. \ref{fig:bubble_growth_dominated}\&\ref{fig:bubble_growth_regime}), which had an initial, artificial relative viscosity of 0.1, towards nearly complete vesiculation. The asymmetry of the bubble growth is controlled by $\mathrm{Pe}$. The bubble overpressure, radius, and vesicularity, do not vary significantly across the suspension, and the total pressure in the interior of the melt droplet remains modest and shows a characteristic 1/r shape as expected for uniform volume expansion. The velocity field is largest at the drop margin which must expand faster to accommodate the interior vesiculation (Fig. \ref{fig:bubble_growth_regime}D-E).  \par 

\begin{figure}
\begin{center}
\includegraphics[width=4 in]{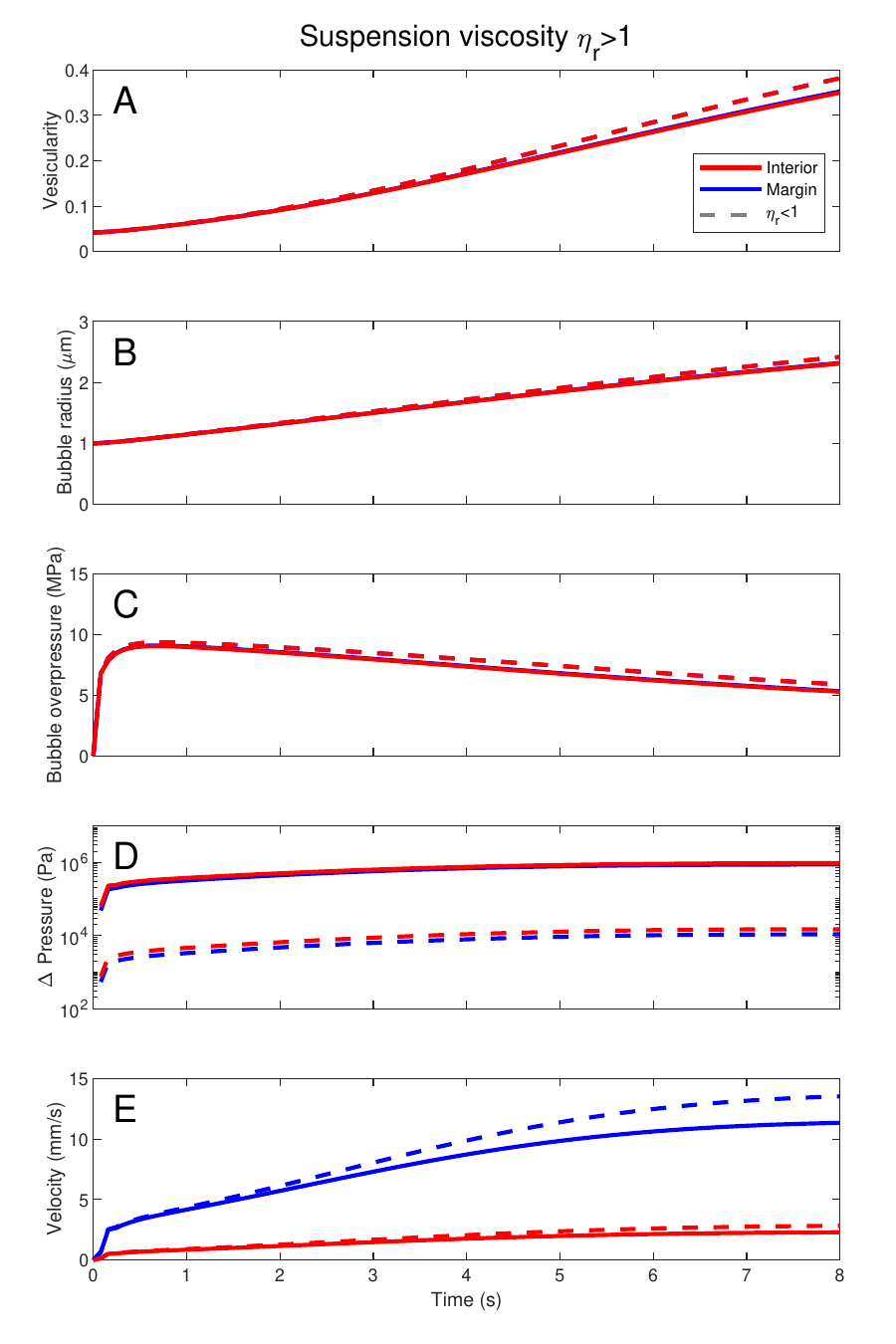}\\
\caption{Simulation results for the isothermal, closed-system, suspension viscosity-limited regime showing A) vesicularity, B) bubble radius, C) bubble overpressure, D) pressure above atmospheric, and E) velocity through time in both the interior (blue) and margin (red) of a spherical droplet, compared to the $\eta_r<1$ case (dashed).}
\label{fig:viscosity_limited}
\end{center}
\end{figure}

Conversely, the presence of crystals, or bubbles in the low $\mathrm{Cc}$ regime, results in a relative viscosity that is larger than unity, suggesting that viscous resistance provided by the deforming suspension can be rate-limiting, especially for viscosity-limited bubble growth. To force a viscosity-limited regime, We repeat the simulation in Fig. \ref{fig:bubble_growth_dominated}, with a relative suspension viscosity of 10 (appropriate for about 40 vol\% crystals) and a bubble number density of 10$^{16}$ 1/m$^3$, higher than the previous value of 10$^{11}$ 1/m$^3$. This suspension viscosity-dominated regime produces a larger pressure in the suspension, especially in the interior of the magma parcel (Fig. \ref{fig:viscosity_limited}), but retains the same characteristic shape as the low suspension-viscosity regime. The development of several hundreds of kPa of pressure suppresses vesiculation under these conditions and bubble over-pressure remains nearly constant at a few MPa. At increasing magma droplet or parcel size, and increasing melt viscosity, viscous resistance far from the individual bubble becomes more important. \par 


\subsection{Conduit-wall friction}
\begin{figure}
\begin{center}
\includegraphics[width=\textwidth]{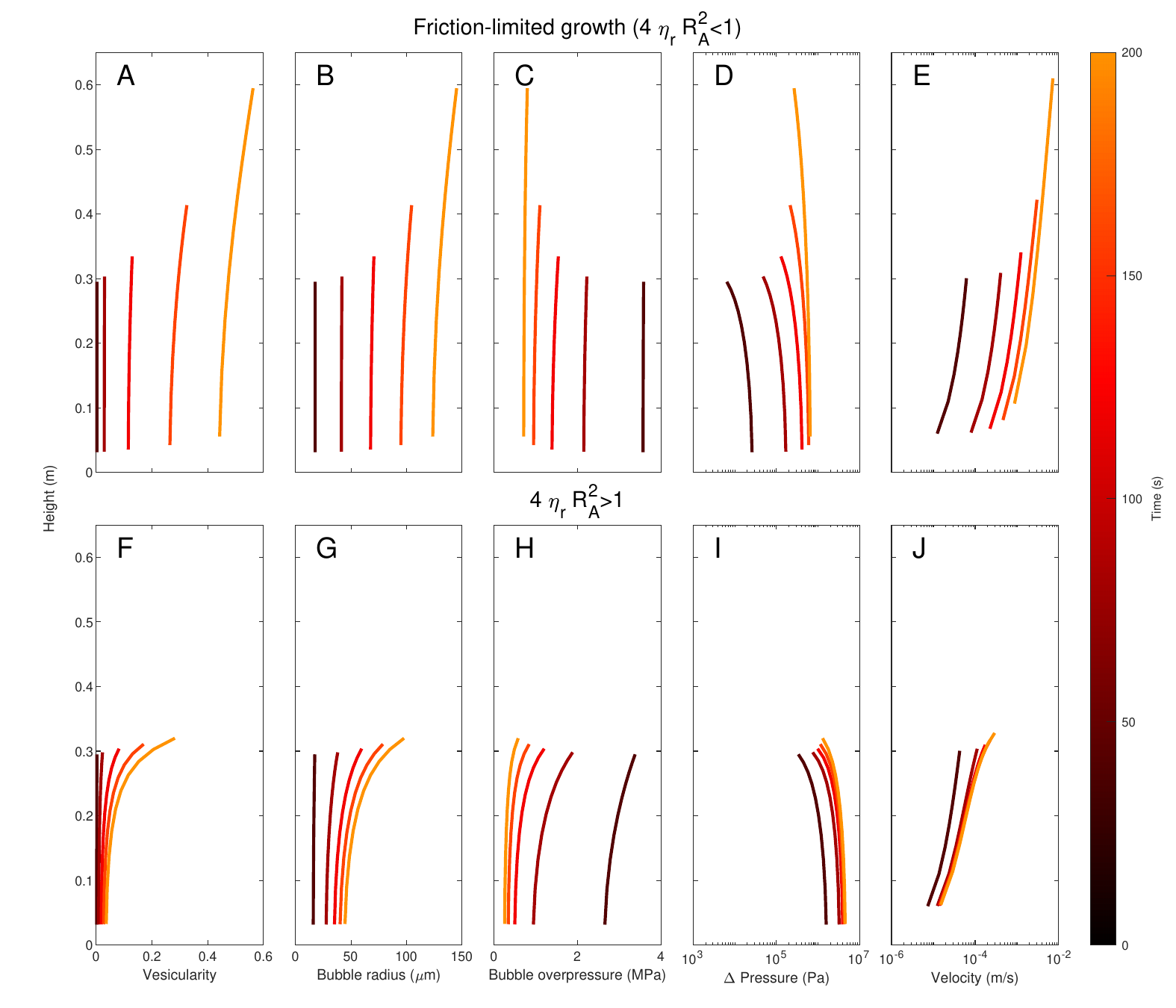}\\
\caption{Simulation results for the isothermal, closed-system, friction-limited regime in A-E) wide (R=1 m) and F-J) narrow (R=2.5 cm) cylindrical conduits, showing A\&F) vesicularity, B\&G) bubble radius, C\&H) bubble overpressure, D\&I) pressure above atmospheric, and E\&J) velocity, with color indicating increasing time from black to red.}
\label{fig:small_conduit}
\end{center}
\end{figure}

In the case of magma expanding in a confined cylindrical conduit, the friction provided by the conduit walls can further impede bubble expansion. In the limit where the conduit is wide, the bubbles are free to expand (in their usual diffusion- or viscosity-limited regime) and the total pressure remains relatively low, but now with a parabolic shape rather than the 1/r that defines the spherical geometry (Fig. \ref{fig:small_conduit}A-D). If the conduit is narrow, expansion can only occur near the free surface (initially L$\lesssim$R/2), below this level the total pressure and the bubble pressure are of the same order which prevents bubble expansion, even as water continues to diffuse into the bubbles. The total pressure in the suspension increases with time and ultimately must be accommodated by at least one of two processes: 1) the pressure within the bubbles becomes large enough to increase the solubility of water in the melt phase until the bubbles stop growing, or 2) the suspension must rupture. Additionally, the cylindrical geometry produces greater shear strains that can move the bubbles into the high capillary number regime, which further decreases the suspension viscosity only in the mobile upper conduit, which creates positive feedbacks in the ease of flow which localizes at the free surface. \par

\subsection{Thermal quenching and high-viscosity regions}
Temperature plays an important role in bubble growth: water diffusivity decreases and viscosity increases with temperature, which means cooling systems hinder and eventually arrest bubble growth. Given that the thermal diffusivity of magmas is typically higher than the water diffusivity ($\mathrm{R_D}<<1$), we see that the critical scaling for the competition between cooling and diffusive bubble growth depends on both the ratio of diffusivities and the ratio of length scales, where cooling is proportional to the surface area. The result is that cooling quenches bubble growth more effectively at smaller system length scales (Fig. \ref{fig:thermal_quenching_regime}), consistent with previous works \citep{Hort2000,Benage2014,Wright2007}. \par

\begin{figure}
\begin{center}
\includegraphics[width=4 in]{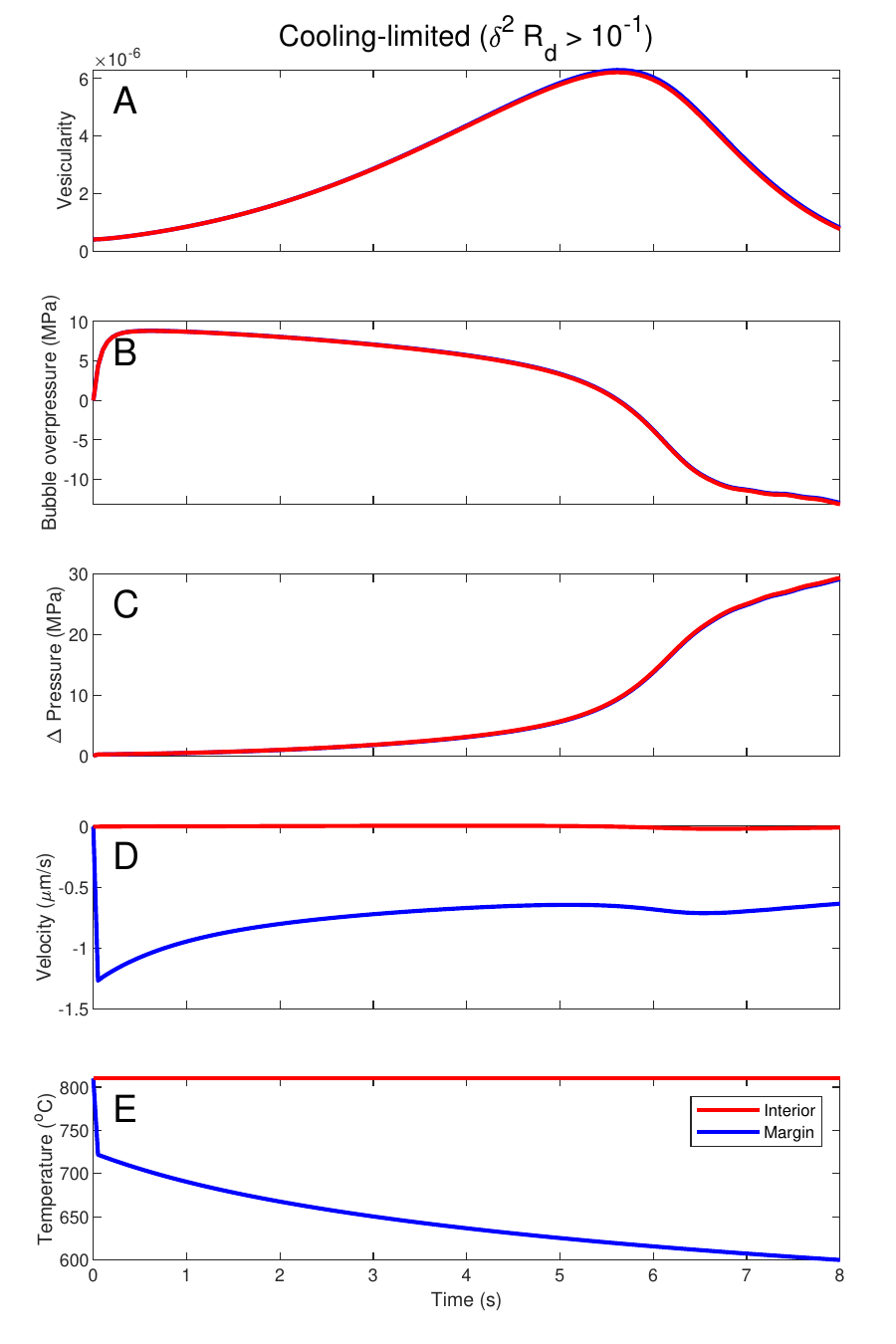}\\
\caption{Simulation results for a cooling melt fragment, showing A) vesicularity, B) bubble overpressure, C) pressure above atmospheric, D) velocity, and E) temperature, with color indicating time the clast margin (blue) and interior (red).}
\label{fig:thermal_quench}
\end{center}
\end{figure}

The quenching of a temperature-controlled rind can have significant effects on the overall bubble growth. 
In this rind, cooling serves to suppress bubble growth by four mechanisms, 1) increases the solubility of water in the lava, 2) decreases the diffusivity of water, 3) reduces the density of vapor in the bubbles, which in turn decreases bubble overpressure, and 4) increases the viscosity of the melt resisting bubble expansion \citep{Kaminski1997}. In many cases, this results in very high bubble overpressure inside the melt which either inhibits bubble growth or contributes to secondary fragmentation \citep{Namiki2021}. In the example simulation shown in Fig. \ref{fig:thermal_quench}, the hoop stress in the cooled rind of a 5 cm radius fragment reaches $>$4 GPa, which would exceed the tensile strength of the melt at 500-600 $^\circ$C \citep{Ahmed2022}. Eventually, the pressure in the melt phase not only inhibits melt growth, but the continued cooling-shrinkage of the rind continues to increase the pressure until the the vesicles resorb completely.

\section{Discussion}
\subsection{Volcanic regimes}
We demonstrate a number of regimes in which silicate melt suspensions are able to exchange momentum, and heat with their surroundings whose dynamics are sensitive to the characteristic length scales of both the melt shell thickness and the melt parcel. Given the complex nature of volcanic systems, the characteristic length scales for magmatic suspensions can vary from micrometers to decimeters in natural pyroclasts up to meters and kilometers for volcanic conduits, magma chambers, and lava flows and domes. Controlled laboratory experiments also span from micrometers to meters. This flexible numerical model allows for exploration across a wide range of both absolute spatial scales and potential regimes arising from the multiscale nature of the problem, and allows for self-consistent interpretation and application of laboratory experiments to natural-scale problems. \par 

In this first demonstration of the model capabilities, we evaluate the effect of having a relative viscosity of the suspension that is far from unity. Previous work on the effective rheology of suspensions has highlighted the importance of the relative size of the bubbles and crystals in which the suspension of the smaller phase serves as the suspending medium for the larger phase \citep[e.g.,][]{Phan-Thien1997,Truby2015,Birnbaum2021}. The bubble growth problem at the microscale has traditionally neglected the effect of crystals or other bubbles which may be in the suspension, and used only the melt viscosity. Our results indicate that the bubble growth is impeded by both the melt and suspension viscosity, 
even when assuming that the crystals are much larger than the bubbles. Although we do not consider the local effects of bubbles and crystals in near contact that are doubtless important to the development of individual bubbles, in aggregate we may expect that the vesiculation of the total suspension must account for the displacement of crystals or neighboring bubbles. Depending on the geometry, the resistance to expansion must also incorporate the flow of higher viscosity regions, which require much higher bubble pressures under otherwise similar conditions. \par 

\subsection{Model limitations}
Although this model presents a significant step forward in coupling the complex processes of bubble evolution with suspension-scale flow, we nonetheless must make several simplifications. A primary challenge is that the bubble-scale model assumes a radially-symmetric spherical bubble that is isolated from its neighboring bubbles or crystals, without allowing for bubble stretching to change the surface area of the bubble available for diffusion, potential shielding of water by crystal- bubble interactions, and bubble-bubble interactions including coalescence or Ostwald ripening, and indeed the nucleation of those bubbles. In theory, the modular approach of the model would allow future extensions to address some of these issues, especially nucleation and coalescence at the population level. The shielding by a large volume fraction of crystals could be incorporated by modifying either the thickness of the melt film or by decreasing the effective diffusivity of the melt depending on the relative size of the crystals and bubbles. Relaxing the assumptions of symmetry at both the bubble and suspension levels is conceptually straightforward, but adds computational expense. 

\subsection{Model extensions}
In \citet{Colombier2026}, we extend the suspension-level dynamics model to two-dimensional cylindrical geometries in which we resolve radial, but not circumferential variation, but we stopped short of0 the fully-coupled model. In that work, the vesiculation history was directly observed via in situ 4D X-Ray computed microtomography which provided inputs of the volume source due to vesiculation into the flow model without resolving the bubble-scale dynamics. Given the modular nature of the software and the existing capability for 2D suspension-scale flow, the model could be run directly in 2D with no additional modifications, but this would also add computational expense and could potentially produce strong feedbacks like shear localization that may destabilize the problem, and as such has not been rigorously tested. \par 

Additionally, we want to highlight that although the constitutive relationships defined within the model are specific to silicate melts, they can easily be adapted to other materials and should be extended to include temporal evolution of those properties in addition to the state variables which could allow, for example, simulation of the development of crust in bread for which there is high-quality experimental data available \citep[e.g.,][]{Babin2006,Zhang2007,Wagner2008,Vanin2009,Grenier2010,Lucas2015,Nicolas2017}, and has many similar underlying processes \citep{deCindio1995,Zhang2006}, or for the flow of polyurethane flows for which one-directional \citep{Geier2014} or partially coupled \citep[e.g.,][]{Ferkl2016,Karolius2017,Rusche2019} multiscale models exist. \par 

\section{Conclusions}
We present a generalized, extensible model for coupled bubble-scale growth and suspensions-scale flow and exchange of mass, momentum, and heat with the surroundings of a finite volume of magma. In this work, we demonstrate the capability of the model to capture the effects of spatial or spatial-scale variability in viscosity and friction effects. In contrast to previous works, we find that the characteristic length scale controlling bubble growth via diffusion is the melt film thickness, rather than the bubble radius which substantially changes the prediction for where regime transitions in bubble growth should occur in magmatic systems. Additionally, we find that bubble growth must account for the contribution of rigid crystals to the suspension viscosity, even when those crystals are large compared to the bubbles. We provide a rigorous treatment of the cooling rind on a pyroclast to simulate the suppression of vesiculation and eventually produce resorption of the vesicles, which should in future be combined with rupture criteria to simulate bread-crust textures. Numerical results documenting the effects of the loss of volatiles to the surroundings through diffusive outgassing and gas percolation will be addressed in future works. This model will improve our ability to interpret laboratory-scale experiments and scale-up to natural processes. 

\section{Funding statement}
We acknowledge funding support from the European Research Council (ERC) under award Magma Outgassing During Eruptions and Geothermal Exploration (MODERATE No.101001065).

\section{Competing interests}
The authors have no competing interests to declare. 

\section{Data availability}
All software accompanying this work can be found at \url{https://github.com/JanineBirnbaum18/MultiscaleBubbles}.

\section{Author contributions}
JB: Conceptualization, methodology, software, formal analysis, investigation, writing - original draft, visualization; FBW: conceptualization, formal analysis, writing - review \& editing;  AL: conceptualization, writing - review \& editing; JEK: conceptualization, writing - review \& editing, supervision, funding acquisition; YL: conceptualization, writing - review \& editing, supervision, funding acquisition.

\bibliography{references}

@article{Bonney1918,
    title = {{"Bread-crust" volcanic bombs [Letters to the editor]}},
    year = {1918},
    journal = {Nature},
    author = {Bonney, T. G.},
    number = {2532},
    month = {5},
    pages = {184},
    volume = {101},
    doi = {10.1038/101184a0}
}

@article{Duffield1972,
    title = {{A naturally occurring model of global plate tectonics}},
    year = {1972},
    journal = {Journal of Geophysical Research},
    author = {Duffield, Wendell A.},
    number = {14},
    month = {5},
    pages = {2543--2555},
    volume = {77},
    publisher = {American Geophysical Union (AGU)},
    doi = {10.1029/jb077i014p02543}
}

@article{Huber2014,
    title = {{A new bubble dynamics model to study bubble growth, deformation, and coalescence}},
    year = {2014},
    journal = {Journal of Geophysical Research: Solid Earth},
    author = {Huber, C. and Su, Y. and Nguyen, C. T. and Parmigiani, A. and Gonnermann, H. M. and Dufek, J.},
    number = {1},
    month = {1},
    pages = {216--239},
    volume = {119},
    publisher = {Blackwell Publishing Ltd},
    doi = {10.1002/2013JB010419},
    issn = {21699356}
}

@article{Yamada2005,
    title = {{A new theory of bubble formation in magma}},
    year = {2005},
    journal = {Journal of Geophysical Research: Solid Earth},
    author = {Yamada, Kou and Tanaka, Hidekazu and Nakazawa, Kiyoshi and Emori, Hiroyuki},
    number = {B02203},
    month = {2},
    pages = {1--17},
    volume = {110},
    publisher = {Blackwell Publishing Ltd},
    doi = {10.1029/2004JB003113},
    issn = {21699356}
}

@article{Sinclair2015,
    title = {{A review of simple formulae for elastic hoop stresses in cylindrical and spherical pressure vessels: WHAT can be used when}},
    year = {2015},
    journal = {International Journal of Pressure Vessels and Piping},
    author = {Sinclair, G. B. and Helms, J. E.},
    month = {4},
    pages = {1--7},
    volume = {128},
    publisher = {Elsevier Ltd},
    doi = {10.1016/j.ijpvp.2015.01.006},
    issn = {03080161}
}

@article{Mancini2016,
    title = {{An expansion-coalescence model to track gas bubble populations in magmas}},
    year = {2016},
    journal = {Journal of Volcanology and Geothermal Research},
    author = {Mancini, Simona and Forestier-Coste, Louis and Burgisser, Alain and James, François and Castro, Jonathan},
    month = {3},
    pages = {44--58},
    volume = {313},
    publisher = {Elsevier B.V.},
    doi = {10.1016/j.jvolgeores.2016.01.016},
    issn = {03770273}
}

@article{DelGaudio2009,
    title = {{An experimental study on the pressure dependence of viscosity in silicate melts}},
    year = {2009},
    journal = {Journal of Chemical Physics},
    author = {Del Gaudio, Piero and Behrens, Harald},
    number = {4},
    month = {7},
    pages = {1--14},
    volume = {131},
    doi = {10.1063/1.3169455},
    issn = {00219606}
}

@article{Coumans2020b,
    title = {{An experimentally validated numerical model for bubble growth in magma}},
    year = {2020},
    journal = {Journal of Volcanology and Geothermal Research},
    author = {Coumans, J. P. and Llewellin, E. W. and Wadsworth, F. B. and Humphreys, M. C.S. and Mathias, S. A. and Yelverton, B. M. and Gardner, J. E.},
    month = {9},
    volume = {402},
    publisher = {Elsevier B.V.},
    doi = {10.1016/j.jvolgeores.2020.107002},
    issn = {03770273}
}

@article{Toramaru2006,
    title = {{BND (bubble number density) decompression rate meter for explosive volcanic eruptions}},
    year = {2006},
    journal = {Journal of Volcanology and Geothermal Research},
    author = {Toramaru, A.},
    number = {3-4},
    pages = {303--316},
    volume = {154},
    doi = {10.1016/j.jvolgeores.2006.03.027},
    issn = {03770273}
}

@article{Wright2007,
    title = {{Breadcrust bombs as indicators of Vulcanian eruption dynamics at Guagua Pichincha volcano, Ecuador}},
    year = {2007},
    journal = {Bulletin of Volcanology},
    author = {Wright, Heather M.N. and Cashman, Katharine V. and Rosi, Mauro and Cioni, Raffaello},
    number = {3},
    month = {1},
    pages = {281--300},
    volume = {69},
    doi = {10.1007/s00445-006-0073-6},
    issn = {02588900}
}

@article{Namiki2021,
    title = {{Brittle fragmentation by rapid gas separation in a Hawaiian fountain}},
    year = {2021},
    journal = {Nature Geoscience},
    author = {Namiki, Atsuko and Patrick, Matthew R. and Manga, Michael and Houghton, Bruce F.},
    number = {4},
    month = {4},
    pages = {242--247},
    volume = {14},
    publisher = {Nature Research},
    doi = {10.1038/s41561-021-00709-0},
    issn = {17520908}
}

@article{Navon1998,
    title = {{Bubble growth in highly viscous melts: theory, experiments, and autoexplosivity of dome lavas}},
    year = {1998},
    journal = {Earth and Planetary Science Letters},
    author = {Navon, Oded and Chekhmir, Anatoly and Lyakhovsky, Vladimir},
    number = {3-4},
    pages = {763--776},
    volume = {160},
    doi = {https://doi.org/10.1016/S0012-821X(98)00126-5}
}

@article{Lyakhovsky1996,
    title = {{Bubble growth in rhyolitic melts: experimental and numerical investigation}},
    year = {1996},
    journal = {Bulletin of Volcanology},
    author = {Lyakhovsky, Vladimir and Hurwitz, Shaul and Navon, Oded},
    pages = {19--32},
    volume = {58},
    publisher = {Springer-Verlag}
}

@article{Dietterich2022,
    title = {{Can lava flow like water? Assessing applications of critical flow theory to channelized basaltic lava flows}},
    year = {2022},
    journal = {Journal of Geophysical Research: Earth Surface},
    author = {Dietterich, H. R. and Grant, G. E. and Fasth, B. and Major, J. J. and Cashman, K. V.},
    number = {9},
    month = {5},
    pages = {1--26},
    volume = {127},
    publisher = {Informa UK Limited},
    doi = {10.1029/2022JF006666}
}

@article{Lister1998,
    title = {{Capillary breakup of a viscous thread surrounded by another viscous fluid}},
    year = {1998},
    journal = {Physics of Fluids},
    author = {Lister, John R and Stone, Howard A},
    number = {11},
    month = {11},
    pages = {2758--2764},
    volume = {10},
    url = {http://ojps.aip.org/phf/phfcr.jsp}
}

@article{Grenier2010,
    title = {{Combining local pressure and temperature measurements during bread baking: insights into crust properties and alveolar structure of crumb}},
    year = {2010},
    journal = {Journal of Cereal Science},
    author = {Grenier, D. and Le Ray, D. and Lucas, T.},
    number = {1},
    month = {7},
    pages = {1--8},
    volume = {52},
    doi = {10.1016/j.jcs.2009.09.009},
    issn = {07335210}
}

@article{Malfait2011,
    title = {{Compositional dependent compressibility of dissolved water in silicate glasses}},
    year = {2011},
    journal = {American Mineralogist},
    author = {Malfait, Wim J. and Sanchez-Valle, Carmen and Ardia, Paola and M{\'{e}}dard, Etienne and Lerch, Philippe},
    number = {8-9},
    pages = {1402--1409},
    volume = {96},
    publisher = {Walter de Gruyter GmbH},
    doi = {10.2138/am.2011.3718},
    issn = {19453027}
}

@article{Hort2000,
    title = {{Constraints on cooling and degassing of pumice during Plinian volcanic eruptions based on model calculations}},
    year = {2000},
    journal = {Journal of Geophysical Research: Solid Earth},
    author = {Hort, M. and Gardner, J.},
    number = {B11},
    month = {11},
    pages = {25981--26001},
    volume = {105},
    publisher = {Blackwell Publishing Ltd},
    doi = {10.1029/2000jb900186},
    issn = {21699356}
}

@article{Cassidy2018,
    title = {{Controls on explosive-effusive volcanic eruption styles}},
    year = {2018},
    journal = {Nature Communications},
    author = {Cassidy, Mike and Manga, Michael and Cashman, Kathy and Bachmann, Olivier},
    number = {2839},
    month = {7},
    pages = {1--16},
    volume = {9},
    publisher = {Nature Publishing Group},
    doi = {10.1038/s41467-018-05293-3},
    issn = {20411723},
    pmid = {30026543}
}

@article{Geier2014,
    title = {{Coupled Macro and Micro-scale Modeling of Polyurethane Foaming Processes}},
    year = {2014},
    journal = {The Journal of Computational Muliphase Flows},
    author = {Geier, Stephanie and Piesche, Manfred},
    number = {4},
    month = {12},
    pages = {377--390},
    volume = {6},
    doi = {10.1260/1757-482X.6.4.377}
}

@article{Vanin2009,
    title = {{Crust formation and its role during bread baking}},
    year = {2009},
    journal = {Trends in Food Science and Technology},
    author = {Vanin, F. M. and Lucas, T. and Trystram, G.},
    number = {8},
    month = {8},
    pages = {333--343},
    volume = {20},
    doi = {10.1016/j.tifs.2009.04.001},
    issn = {09242244}
}

@article{Phan-Thien1997,
    title = {{Differential multiphase models for polydispersed suspensions and particulate solids}},
    year = {1997},
    journal = {Journal of Non-Newtonian Fluid Mechanics},
    author = {Phan-Thien, N. and Pham, D. C.},
    pages = {305--318},
    volume = {72},
    doi = {10.1016/S0377-0257(97)90002-1},
    issn = {03770257}
}

@article{Zhang2010,
    title = {{Diffusion of H, C, and O components in silicate melts}},
    year = {2010},
    journal = {Reviews in Mineralogy and Geochemistry},
    author = {Zhang, Youxue and Ni, Huaiwei},
    pages = {171--225},
    volume = {72},
    isbn = {9780939950867},
    doi = {10.2138/rmg.2010.72.5},
    issn = {15296466}
}

@article{McIntosh2014,
    title = {{Distribution of dissolved water in magmatic glass records growth and resorption of bubbles}},
    year = {2014},
    journal = {Earth and Planetary Science Letters},
    author = {McIntosh, I. M. and Llewellin, E. W. and Humphreys, M. C.S. and Nichols, A. R.L. and Burgisser, A. and Schipper, C. I. and Larsen, J. F.},
    month = {9},
    pages = {1--11},
    volume = {401},
    publisher = {Elsevier},
    doi = {10.1016/j.epsl.2014.05.037},
    issn = {0012821X}
}

@article{Proussevitch1998,
    title = {{Dynamics and energetics of bubble growth in magmas: Analytical formulation and numerical modeling}},
    year = {1998},
    journal = {Journal of Geophysical Research},
    author = {Proussevitch, A. A. and Sahagian, D. L.},
    number = {B8},
    month = {8},
    pages = {18223--18251},
    volume = {103},
    publisher = {Blackwell Publishing Ltd},
    doi = {10.1029/98jb00906},
    issn = {21699356}
}

@article{Proussevitch1996,
    title = {{Dynamics of coupled diffusive and decompressive bubble growth in magmatic systems}},
    year = {1996},
    journal = {Journal of Geophysical Research: Solid Earth},
    author = {Proussevitch, Alexander A. and Sahagian, Dork L.},
    number = {B8},
    month = {8},
    pages = {17447--17455},
    volume = {101},
    publisher = {American Geophysical Union},
    doi = {10.1029/96jb01342},
    issn = {21699356}
}

@article{Prousevitch1993,
    title = {{Dynamics of diffusive bubble growth in magmas: isothermal case}},
    year = {1993},
    journal = {Journal of Geophysical Research},
    author = {Prousevitch, A. A. and Sahagian, D. L. and Anderson, A. T.},
    number = {B12},
    month = {12},
    pages = {22283--22307},
    volume = {98},
    doi = {10.1029/93jb02027},
    issn = {01480227}
}

@article{Proussevitch1993,
    title = {{Dynamics of diffusive bubble growth in magmas: isothermal case}},
    year = {1993},
    journal = {Journal of Geophysical Research},
    author = {Proussevitch, A. A. and Sahagian, D. L. and Anderson, A. T.},
    number = {B12},
    volume = {98},
    doi = {10.1029/93jb02027},
    issn = {01480227}
}

@article{Melnik2005b,
    title = {{Dynamics of magma flow inside volcanic conduits with bubble overpressure buildup and gas loss through permeable magma}},
    year = {2005},
    journal = {Journal of Volcanology and Geothermal Research},
    author = {Melnik, O. and Barmin, A. A. and Sparks, R. Steve J.},
    number = {1-3},
    month = {5},
    pages = {53--68},
    volume = {143},
    doi = {10.1016/j.jvolgeores.2004.09.010},
    issn = {03770273}
}

@article{Bagdassarov2000,
    title = {{Effect of alkalis, phosphorus, and water on the surface tension of haplogranite melt}},
    year = {2000},
    journal = {American Mineralogist},
    author = {Bagdassarov, Nikolai and Dorfman, Alexander and Dingwell, Donald B},
    number = {1},
    month = {1},
    pages = {33--40},
    volume = {85},
    isbn = {00/00010033{\$}05.0}
}

@article{Ahmed2022,
    title = {{Effect of elevated temperature on rhyolitic rocks’ properties}},
    year = {2022},
    journal = {Materials},
    author = {Ahmed, Haitham M. and Hefni, Mohammed A. and Ahmed, Hussin A.M. and Adewuyi, Sefiu O. and Hassani, Ferri and Sasmitob, Agus P. and Saleem, Hussein A. and Moustafa, Essam B. and Hassan, Gamal S.A.},
    number = {9},
    month = {4},
    pages = {1--24},
    volume = {15},
    publisher = {MDPI},
    doi = {10.3390/ma15093204},
    issn = {19961944}
}

@article{Bouhifd2006,
    title = {{Effect of water on the heat capacity of polymerized aluminosilicate glasses and melts}},
    year = {2006},
    journal = {Geochimica et Cosmochimica Acta},
    author = {Bouhifd, M. Ali and Whittington, Alan and Roux, Jacques and Richet, Pascal},
    number = {3},
    month = {2},
    pages = {711--722},
    volume = {70},
    doi = {10.1016/j.gca.2005.09.012},
    issn = {00167037}
}

@article{Zhang2007,
    title = {{Effects of crust constraints on bread expansion and CO2 release}},
    year = {2007},
    journal = {Journal of Food Engineering},
    author = {Zhang, Lu and Lucas, T. and Doursat, C. and Flick, D. and Wagner, M.},
    number = {4},
    month = {6},
    pages = {1302--1311},
    volume = {80},
    doi = {10.1016/j.jfoodeng.2006.10.008},
    issn = {02608774}
}

@article{Griffiths1993,
    title = {{Effects of surface cooling on the spreading of lava flows and domes}},
    year = {1993},
    journal = {J. Fluid Mech},
    author = {Griffiths, Ross W. and Fink, Jonathan H},
    pages = {661--102},
    volume = {252}
}

@article{Pitzer1994,
    title = {{Equations of state valid continuously from zero to extreme pressures for H2O and CO2}},
    year = {1994},
    journal = {The Journal of Chemical Physics},
    author = {Pitzer, Kenneth S. and Sterner, S. Michael},
    number = {4},
    pages = {3111--3116},
    volume = {101},
    doi = {10.1063/1.467624},
    issn = {00219606}
}

@article{Kaminski1997,
    title = {{Expansion and quenching of vesicular magma fragments in Plinian eruptions}},
    year = {1997},
    journal = {Journal of Geophysical Research: Solid Earth},
    author = {Kaminski, Edouard and Jaupart, Claude},
    number = {B6},
    month = {6},
    pages = {12187--12203},
    volume = {102},
    publisher = {American Geophysical Union},
    doi = {10.1029/97jb00622},
    issn = {21699356}
}

@article{Nicolas2017,
    title = {{Experiment and multiphysic simulation of dough baking by convection, infrared radiation and direct conduction}},
    year = {2017},
    journal = {International Journal of Thermal Sciences},
    author = {Nicolas, V. and Glouannec, P. and Ploteau, J. P. and Salagnac, P. and Jury, V.},
    month = {5},
    pages = {65--78},
    volume = {115},
    publisher = {Elsevier Masson SAS},
    doi = {10.1016/j.ijthermalsci.2017.01.018},
    issn = {12900729}
}

@article{Gonnermann2003,
    title = {{Explosive volcanism may not be an inevitable consequence of magma fragmentation}},
    year = {2003},
    journal = {Nature},
    author = {Gonnermann, Helge M. and Manga, Michael},
    number = {},
    month = {11},
    pages = {432--435},
    volume = {426},
    doi = {10.1038/nature02138},
    issn = {00280836}
}

@article{LaSpina2021,
    title = {{Explosivity of basaltic lava fountains is controlled by magma rheology, ascent rate and outgassing}},
    year = {2021},
    journal = {Earth and Planetary Science Letters},
    author = {La Spina, G. and Arzilli, F. and Llewellin, E. W. and Burton, M. R. and Clarke, A. B. and de' Michieli Vitturi, M. and Polacci, M. and Hartley, M. E. and Di Genova, D. and Mader, H. M.},
    number = {116658},
    month = {1},
    pages = {1--11},
    volume = {553},
    publisher = {Elsevier B.V.},
    doi = {10.1016/j.epsl.2020.116658},
    issn = {0012821X}
}

@article{Babin2006,
    title = {{Fast X-ray tomography analysis of bubble growth and foam setting during breadmaking}},
    year = {2006},
    journal = {Journal of Cereal Science},
    author = {Babin, P. and Della Valle, G. and Chiron, H. and Cloetens, P. and Hoszowska, J. and Pernot, P. and R{\'{e}}guerre, A. L. and Salvo, L. and Dendievel, R.},
    number = {3},
    month = {5},
    pages = {393--397},
    volume = {43},
    doi = {10.1016/j.jcs.2005.12.002},
    issn = {07335210}
}

@article{Sparks1983,
    title = {{Fluid Dynamics in Volcanology (Wager Prize Lecture)}},
    year = {1983},
    journal = {Bulletin of Volcanology},
    author = {Sparks, R S J},
    pages = {323--331},
    volume = {46}
}

@article{Jaupart1991,
    title = {{Gas content, eruption rate and instabilities of eruption regime in silicic volcanoes}},
    year = {1991},
    journal = {Earth and Planetary Science Letters},
    author = {Jaupart, Claude and All{\`{e}}gre, Claude J},
    number = {3-4},
    month = {3},
    pages = {413--429},
    volume = {102},
    doi = {https://doi.org/10.1016/0012-821X(91)90032-D}
}

@incollection{Jaupart1998,
    title = {{Gas loss from magmas through conduit walls during eruption}},
    year = {1998},
    booktitle = {The Physcis of Explosive Volcanic Eruptions},
    author = {Jaupart, C},
    editor = {Gilbert, J. S. and Sparks, R. S. J.},
    number = {145},
    pages = {73--90},
    volume = {145},
    publisher = {Geological Society, London, Special Publications},
    url = {http://sp.lyellcollection.org/},
    doi = {https://doi.org/10.1144/GSL.SP.1996.145.01.05}
}

@article{Crosweller2012,
    title = {{Global database on large magnitude explosive volcanic eruptions (LaMEVE)}},
    year = {2012},
    journal = {Journal of Applied Volcanology},
    author = {Crosweller, Helen Sian and Arora, Baneet and Brown, Sarah Krystyna and Cottrell, Elizabeth and Deligne, Natalia Irma and Guerrero, Natalie Ortiz and Hobbs, Laura and Kiyosugi, Koji and Loughlin, Susan Clare and Lowndes, Jonathan and Nayembil, Martin and Siebert, Lee and Sparks, Robert Stephen John and Takarada, Shinji and Venzke, Edward},
    number = {4},
    month = {9},
    pages = {1--13},
    volume = {1},
    publisher = {SpringerOpen},
    doi = {10.1186/2191-5040-1-4},
    issn = {21915040}
}

@article{Schunke2026,
    title = {{Grain size distribution controls sintering of hydrous pyroclasts}},
    year = {2026},
    journal = {Journal of Volcanology and Geothermal Research},
    author = {Schunke, Julia and Wadsworth, Fabian B. and Kendrick, Jackie E. and Lamur, Anthony and Birnbaum, Janine and Lavall{\'{e}}e, Yan},
    number = {108509},
    month = {2},
    pages = {1--16},
    volume = {470},
    publisher = {Elsevier B.V.},
    doi = {10.1016/j.jvolgeores.2025.108509},
    issn = {03770273}
}

@article{Stebbins1984,
    title = {{Heat capacities and entropies of silicate liquids and glasses}},
    year = {1984},
    journal = {Contributions to Mineralogy and Petrology},
    author = {Stebbins, J F and Carmichael, I S E and Moret, L K},
    pages = {131--148},
    volume = {86},
    doi = {https://doi.org/10.1007/BF00381840}
}

@article{Jones2022,
    title = {{Inflated pyroclasts in proximal fallout deposits reveal abrupt transitions in eruption behaviour}},
    year = {2022},
    journal = {Nature Communications},
    author = {Jones, Thomas J. and Le Moigne, Yannick and Russell, James K. and Williams-Jones, Glyn and Giordano, Daniele and Dingwell, Donald B.},
    number = {2832},
    month = {5},
    pages = {1--12},
    volume = {13},
    publisher = {Nature Research},
    doi = {10.1038/s41467-022-30501-6},
    issn = {20411723},
    pmid = {35595774}
}

@article{Colombier2026,
    title = {{Inherent duality of vesiculation kinematics revealed through 4D imaging}},
    year = {2026},
    journal = {Journal of Volcanology and Geothermal Research},
    author = {Colombier, Mathieu and C{\'{a}}ceres, Francisco and Birnbaum, Janine and deGraffenried, Rebecca and Lavall{\'{e}}e, Yan and Kendrick, Jackie E. and Scheu, Bettina and Thivet, Simon and Valdivia, Pedro and Ruhekenya, Ruben M. and Schlep{\"{u}}tz, Christian M. and Castro, Jonathan M. and Hess, Kai Uwe and Dingwell, Donald B.},
    number = {108505},
    month = {2},
    pages = {1--13},
    volume = {470},
    publisher = {Elsevier B.V.},
    doi = {10.1016/j.jvolgeores.2025.108505},
    issn = {03770273}
}

@article{Chernov2014,
    title = {{Kinetics of gas bubble nucleation and growth in magmatic melt at its rapid decompression}},
    year = {2014},
    journal = {Physics of Fluids},
    author = {Chernov, A. A. and Kedrinsky, V. K. and Pil'nik, A. A.},
    number = {11},
    month = {11},
    pages = {1--19},
    volume = {26},
    publisher = {American Institute of Physics Inc.},
    doi = {10.1063/1.4900846},
    issn = {10897666}
}

@incollection{Burgisser2015,
    title = {{Magma Ascent and Degassing at Shallow Levels}},
    year = {2015},
    booktitle = {The Encyclopedia of Volcanoes},
    author = {Burgisser, Alain and Degruyter, Wim},
    editor = {Sigurdsson, Haraldur and Houghton, Bruce and McNutt, Stephen R. and Rymer, Hazel and Stix, John},
    chapter = {11},
    edition = {2},
    pages = {225--236},
    publisher = {Elsevier},
    isbn = {9780123859389},
    doi = {10.1016/B978-0-12-385938-9.00011-0}
}

@article{Rivalta2008,
    title = {{Magma compressibility and the missing source for some dike intrusions}},
    year = {2008},
    journal = {Geophysical Research Letters},
    author = {Rivalta, Eleonora and Segall, Paul},
    number = {4},
    month = {2},
    volume = {35},
    doi = {10.1029/2007GL032521},
    issn = {00948276}
}

@article{Scheu2022,
    title = {{Magma Fragmentation}},
    year = {2022},
    journal = {Reviews in Mineralogy and Geochemistry},
    author = {Scheu, Bettina and Dingwell, Donald B.},
    pages = {767--800},
    volume = {87},
    publisher = {Mineralogical Society of America},
    doi = {10.2138/rmg.2021.87.16},
    issn = {19432666}
}

@article{Spieler2004,
    title = {{Magma fragmentation speed: An experimental determination}},
    year = {2004},
    journal = {Journal of Volcanology and Geothermal Research},
    author = {Spieler, O. and Dingwell, D. B. and Alidibirov, M.},
    number = {1-3},
    month = {1},
    pages = {109--123},
    volume = {129},
    publisher = {Elsevier},
    doi = {10.1016/S0377-0273(03)00235-X},
    issn = {03770273}
}

@article{Zhang2006,
    title = {{Mathematical modeling of bread baking process}},
    year = {2006},
    journal = {Journal of Food Engineering},
    author = {Zhang, J. and Datta, A. K.},
    number = {1},
    month = {7},
    pages = {78--89},
    volume = {75},
    doi = {10.1016/j.jfoodeng.2005.03.058},
    issn = {02608774}
}

@article{deCindio1995,
    title = {{Mathematical Modelling of Leavened Cereal Goods}},
    year = {1995},
    journal = {Journal of Food Engineering},
    author = {De Cindio, B and Correra, S},
    pages = {379--403},
    volume = {24}
}

@article{Castro2012a,
    title = {{Mechanisms of bubble coalescence in silicic magmas}},
    year = {2012},
    journal = {Bulletin of Volcanology},
    author = {Castro, Jonathan M. and Burgisser, Alain and Schipper, C. Ian and Mancini, Simona},
    number = {10},
    month = {12},
    pages = {2339--2352},
    volume = {74},
    publisher = {Springer Verlag},
    doi = {10.1007/s00445-012-0666-1},
    issn = {14320819}
}

@article{Lucas2015,
    title = {{Modeling of bread baking with a new, multi-scale formulation of evaporation-condensation-diffusion and evidence of compression in the outskirts of the crumb}},
    year = {2015},
    journal = {Journal of Food Engineering},
    author = {Lucas, T. and Doursat, C. and Grenier, D. and Wagner, M. and Trystram, G. and Flick, D.},
    pages = {24--37},
    volume = {149},
    publisher = {Elsevier Ltd},
    doi = {10.1016/j.jfoodeng.2014.07.020},
    issn = {02608774}
}

@article{Wagner2008,
    title = {{MRI evaluation of local expansion in bread crumb during baking}},
    year = {2008},
    journal = {Journal of Cereal Science},
    author = {Wagner, M. and Quellec, S. and Trystram, G. and Lucas, T.},
    number = {1},
    month = {7},
    pages = {213--223},
    volume = {48},
    doi = {10.1016/j.jcs.2007.09.006},
    issn = {07335210}
}

@article{Ferkl2016,
    title = {{Multi-scale modelling of expanding polyurethane foams: Coupling macro- and bubble-scales}},
    year = {2016},
    journal = {Chemical Engineering Science},
    author = {Ferkl, Pavel and Karimi, Mohsen and Marchisio, Daniele L. and Kosek, Juraj},
    month = {7},
    pages = {55--64},
    volume = {148},
    publisher = {Elsevier Ltd},
    doi = {10.1016/j.ces.2016.03.040},
    issn = {00092509}
}

@article{Eichelberger1986,
    title = {{Non-explosive silicic volcanism}},
    year = {1986},
    journal = {Nature},
    author = {Eichelberger, J C and Carrigan, C R and Westrich, H R and Price, R H},
    number = {},
    month = {10},
    pages = {598--602},
    volume = {323},
    publisher = {Princeton University Press},
    doi = {https://doi.org/10.1038/323598a0}
}

@article{Watkins2017,
    title = {{Nonequilibrium degassing, regassing, and vapor fluxing in magmatic feeder systems}},
    year = {2017},
    journal = {Geology},
    author = {Watkins, J. M. and Gardner, J. E. and Befus, K. S.},
    number = {2},
    pages = {183--186},
    volume = {45},
    publisher = {Geological Society of America},
    doi = {10.1130/G38501.1},
    issn = {19432682}
}

@article{Toramaru1995,
    title = {{Numerical study of nucleation and growth of bubbles in viscous magmas}},
    year = {1995},
    journal = {Journal of Geophysical Research},
    author = {Toramaru, A.},
    number = {B2},
    month = {2},
    pages = {1913--1931},
    volume = {100},
    doi = {10.1029/94JB02775},
    issn = {01480227}
}

@article{Thomas1994,
    title = {{On the vesicularity of pumice}},
    year = {1994},
    journal = {Journal of Geophysical Research},
    author = {Thomas, N. and Jaupart, C. and Vergniolle, S.},
    number = {B8},
    month = {8},
    pages = {15633--15644},
    volume = {99},
    doi = {10.1029/94jb00650},
    issn = {01480227}
}

@article{vonAulock2017,
    title = {{Outgassing from open and closed magma foams}},
    year = {2017},
    journal = {Frontiers in Earth Science},
    author = {von Aulock, Felix W. and Kennedy, Ben M. and Maksimenko, Anton and Wadsworth, Fabian B. and Lavall{\'{e}}e, Yan},
    month = {6},
    pages = {1--7},
    volume = {5},
    publisher = {Frontiers Media S.A.},
    doi = {10.3389/feart.2017.00046},
    issn = {22966463}
}

@article{Mueller2005,
    title = {{Permeability and degassing of dome lavas undergoing rapid decompression: An experimental determination}},
    year = {2005},
    journal = {Bulletin of Volcanology},
    author = {Mueller, Sebastian and Melnik, Oleg and Spieler, Oliver and Scheu, Bettina and Dingwell, Donald B.},
    number = {6},
    month = {7},
    pages = {526--538},
    volume = {67},
    doi = {10.1007/s00445-004-0392-4},
    issn = {02588900}
}

@article{Shah1998,
    title = {{Proving of bread dough: Modelling the growth of individual bubbles}},
    year = {1998},
    journal = {Food and Bioproducts Processing},
    author = {Shah, P. and Campbell, G. M. and Mckee, S. L. and Rielly, C. D.},
    number = {2},
    pages = {73--79},
    volume = {76},
    publisher = {Institution of Chemical Engineers},
    doi = {10.1205/096030898531828},
    issn = {09603085}
}

@article{Fink1990,
    title = {{Radial spreading of viscous-gravity currents with solidifying crust}},
    year = {1990},
    journal = {J. Fluid Mech},
    author = {Fink, Jonathan H and Griffiths, Ross W},
    pages = {485--509},
    volume = {221}
}

@article{Lensky2001,
    title = {{Radial variations of melt viscosity around growing bubbles and gas overpressure in vesiculating magmas}},
    year = {2001},
    journal = {Earth and Planetary Science Letters},
    author = {Lensky, Nadav G and Lyakhovsky, Vladimir and Navon, Oded},
    pages = {1--6},
    volume = {186},
    url = {www.elsevier.com/locate/epsl}
}

@article{Pal2003,
    title = {{Rheological behavior of bubble-bearing magmas}},
    year = {2003},
    journal = {Earth and Planetary Science Letters},
    author = {Pal, Rajinder},
    number = {1-4},
    pages = {165--179},
    volume = {207},
    doi = {10.1016/S0012-821X(02)01104-4},
    issn = {0012821X}
}

@article{Birnbaum2021,
    title = {{Rheology of three-phase lava analogues determined via dam-break experiments}},
    year = {2021},
    journal = {Proceedings of the Royal Society A},
    author = {Birnbaum, Janine and Lev, Einat and Llewellin, Edward W},
    number = {2254},
    month = {10},
    pages = {1--16},
    volume = {477},
    doi = {10.1098/rspa.2021.0394},
    pmid = {PMC9097490}
}

@inproceedings{Karolius2017,
    title = {{Sequential Multi-Scale Modelling Concepts Applied to the Polyurethane Foaming Process}},
    year = {2017},
    booktitle = {Proceedings of the 27th European Symposium on Computer Aided Process Engineering},
    author = {Karolius, Sigve and Preisig, Heinz A. and Rusche, Henrik},
    editor = {Espu{\~{n}}a, Antonio and Graells, Moisès and Puigjaner, Luis},
    month = {10},
    pages = {487--492},
    volume = {40},
    publisher = {Elsevier B.V.},
    address = {Barcelona, Spain},
    doi = {10.1016/B978-0-444-63965-3.50083-0},
    issn = {15707946}
}

@article{Vasseur2023,
    title = {{Shear thinning and brittle failure in crystal-bearing magmas arise from local non-Newtonian effects in the melt}},
    year = {2023},
    journal = {Earth and Planetary Science Letters},
    author = {Vasseur, Jérémie and Wadsworth, Fabian B. and Dingwell, Donald B.},
    number = {117988},
    month = {2},
    pages = {1--10},
    volume = {603},
    publisher = {Elsevier B.V.},
    doi = {10.1016/j.epsl.2023.117988},
    issn = {0012821X}
}

@article{Birnbaum2026a,
    title = {{Shear-enhanced dynamic permeability development of magma vesiculating in cylindrical conduits}},
    year = {2026},
    journal = {Scientific Reports},
    author = {Birnbaum, J. and Schauroth, J. and Weaver, J. and Kendrick, J. E. and Lamur, A. and Lavall{\'{e}}e, Y.},
    number = {9838},
    month = {3},
    pages = {1--16},
    volume = {16},
    doi = {10.1038/s41598-026-43344-8}
}

@incollection{Rusche2019,
    title = {{Simulating polyurethane foams using the modena multi-scale simulation framework}},
    year = {2019},
    booktitle = {OpenFOAM - Selected Papers of the 11th Workshop},
    author = {Rusche, Henrik and Karimi, Mohsen and Ferkl, Pavel and Karolius, Sigve},
    editor = {N{\'{o}}brega, J. and Jasak, H.},
    month = {1},
    pages = {401--417},
    publisher = {Springer, Cham},
    isbn = {9783319608457},
    doi = {10.1007/978-3-319-60846-4{\_}29}
}

@article{Griffiths1992,
    title = {{Solidification and morphology of submarine lavas: a dependence on extrusion rate}},
    year = {1992},
    journal = {Journal of Geophysical Research},
    author = {Griffiths, R. W. and Fink, J. H.},
    number = {B13},
    month = {12},
    pages = {19729--19737},
    volume = {97},
    doi = {10.1029/92jb01594},
    issn = {01480227}
}

@article{Liu2005,
    title = {{Solubility of H2O in rhyolitic melts at low pressures and a new empirical model for mixed H2O-CO2 solubility in rhyolitic melts}},
    year = {2005},
    journal = {Journal of Volcanology and Geothermal Research},
    author = {Liu, Yang and Zhang, Youxue and Behrens, Harald},
    month = {5},
    pages = {219--235},
    volume = {143},
    doi = {10.1016/j.jvolgeores.2004.09.019},
    issn = {03770273}
}

@article{Dingwell1989,
    title = {{Structural Relaxation in Silicate Melts and Non-Newtonian Melt Rheology in Geologic Processes}},
    year = {1989},
    journal = {Physics of Chemistry and Minerals},
    author = {Dingwell, Donald B and Webb, Sharon L},
    pages = {508--516},
    volume = {16}
}

@article{Fink1992,
    title = {{Textural constraints on effusive silicic volcanism: beyond the permeable foam model}},
    year = {1992},
    journal = {Journal of Geophysical Research},
    author = {Fink, J. H. and Anderson, S. W. and Manley, C. R.},
    number = {B6},
    month = {6},
    pages = {9073--9083},
    volume = {97},
    doi = {10.1029/92JB00416},
    issn = {01480227}
}

@article{Llewellin2002a,
    title = {{The constitutive equation and flow dynamics of bubbly magmas}},
    year = {2002},
    journal = {Geophysical Research Letters},
    author = {Llewellin, E W and Mader, H M and Wilson, S D R},
    number = {24},
    volume = {29},
    doi = {10.1029/2002gl015697},
    issn = {00948276}
}

@article{Sparks1978,
    title = {{The dynamics of bubble formation and growth in magmas: A review and analysis}},
    year = {1978},
    journal = {Journal of Volcanology and Geothermal Research},
    author = {Sparks, R. S.J.},
    pages = {1--37},
    volume = {3},
    isbn = {0377-0273},
    doi = {10.1016/0377-0273(78)90002-1},
    issn = {03770273}
}

@article{Griffiths2000,
    title = {{The dynamics of lava flows}},
    year = {2000},
    journal = {Annu. Rev. Fluid Mech},
    author = {Griffiths, R W},
    pages = {477--518},
    volume = {32},
    url = {www.annualreviews.org}
}

@article{Degruyter2012,
    title = {{The effects of outgassing on the transition between effusive and explosive silicic eruptions}},
    year = {2012},
    journal = {Earth and Planetary Science Letters},
    author = {Degruyter, W. and Bachmann, O. and Burgisser, A. and Manga, M.},
    month = {10},
    pages = {161--170},
    volume = {349-350},
    doi = {10.1016/j.epsl.2012.06.056},
    issn = {0012821X}
}

@incollection{Jaupart1992,
    title = {{The eruption and spreading of lava}},
    year = {1992},
    booktitle = {Chaotic Processes in the Geological Sciences},
    author = {Jaupart, Claude},
    editor = {Yuen, David A},
    pages = {175--203},
    volume = {41},
    publisher = {Springer-Verlag New York, Inc}
}

@article{Spieler2004b,
    title = {{The fragmentation threshold of pyroclastic rocks}},
    year = {2004},
    journal = {Earth and Planetary Science Letters},
    author = {Spieler, Oliver and Kennedy, Ben and Kueppers, Ulrich and Dingwell, Donald B. and Scheu, Bettina and Taddeucci, Jacopo},
    number = {1-2},
    month = {9},
    pages = {139--148},
    volume = {226},
    doi = {10.1016/j.epsl.2004.07.016},
    issn = {0012821X}
}

@article{Rougier2018,
    title = {{The global magnitude–frequency relationship for large explosive volcanic eruptions}},
    year = {2018},
    journal = {Earth and Planetary Science Letters},
    author = {Rougier, Jonathan and Sparks, R. Stephen J. and Cashman, Katharine V. and Brown, Sarah K.},
    month = {1},
    pages = {621--629},
    volume = {482},
    publisher = {Elsevier B.V.},
    doi = {10.1016/j.epsl.2017.11.015},
    issn = {0012821X}
}

@article{Webb1990,
    title = {{The Onset of Non-Newtonian Rheology of Silicate Melts: A Fiber Elongation Study}},
    year = {1990},
    journal = {Physics of Chemistry and Minerals},
    author = {Webb, Sharon L and Dingwell, Donald B},
    pages = {125--132},
    volume = {17},
    publisher = {Springer-Verlag}
}

@article{Llewellin2002b,
    title = {{The rheology of a bubbly liquid}},
    year = {2002},
    journal = {Proceedings of the Royal Society A: Mathematical, Physical and Engineering Sciences},
    author = {Llewellin, E. W. and Mader, H. M. and Wilson, S. D.R.},
    number = {2020},
    pages = {987--1016},
    volume = {458},
    doi = {10.1098/rspa.2001.0924},
    issn = {13645021}
}

@article{Truby2015,
    title = {{The rheology of three-phase suspensions at low bubble capillary number}},
    year = {2015},
    journal = {Proceedings of the Royal Society A: Mathematical, Physical and Engineering Sciences},
    author = {Truby, J M and Mueller, S P and Llewellin, E W and Mader, H M},
    number = {2173},
    volume = {471},
    doi = {10.1098/rspa.2014.0557},
    issn = {14712946}
}

@article{Mader2013,
    title = {{The rheology of two-phase magmas : A review and analysis}},
    year = {2013},
    journal = {Journal of Volcanology and Geothermal Research},
    author = {Mader, H M and Llewellin, E W and Mueller, S P},
    pages = {135--158},
    volume = {257},
    publisher = {Elsevier B.V.},
    url = {http://dx.doi.org/10.1016/j.jvolgeores.2013.02.014},
    doi = {10.1016/j.jvolgeores.2013.02.014},
    issn = {0377-0273}
}

@article{Bagdassarov1994,
    title = {{Thermal properties of vesicular rhyolite}},
    year = {1994},
    journal = {Journal of Volcanology and Geothermal Research},
    author = {Bagdassarov, N and Dingwell, D},
    pages = {179--191},
    volume = {60},
    doi = {https://doi.org/10.1016/0377-0273(94)90067-1}
}

@article{Woods1994,
    title = {{Transitions between explosive and effusive eruptions of silicic magmas}},
    year = {1994},
    journal = {Nature},
    author = {Woods, Andrew W. and Koyaguchi, Takehiro},
    number = {},
    pages = {641--644},
    volume = {370},
    publisher = {Wiley},
    doi = {https://doi.org/10.1038/370641a0}
}

@article{Benage2014,
    title = {{Tying textures of breadcrust bombs to their transport regime and cooling history}},
    year = {2014},
    journal = {Journal of Volcanology and Geothermal Research},
    author = {Benage, Mary C. and Dufek, Josef and Degruyter, Wim and Geist, Dennis and Harpp, Karen and Rader, Erika},
    month = {3},
    pages = {92--107},
    volume = {274},
    publisher = {Elsevier B.V.},
    doi = {10.1016/j.jvolgeores.2014.02.005},
    issn = {03770273}
}

@article{Weaver2022,
    title = {{Vesiculation and densification of pyroclasts: A clast-size dependent competition between bubble growth and diffusive outgassing}},
    year = {2022},
    journal = {Journal of Volcanology and Geothermal Research},
    author = {Weaver, Joshua and Lavall{\'{e}}e, Yan and Ashraf, Maliha and Kendrick, Jackie E. and Lamur, Anthony and Schauroth, Jenny and Wadsworth, Fabian B.},
    month = {8},
    volume = {428},
    publisher = {Elsevier B.V.},
    doi = {10.1016/j.jvolgeores.2022.107550},
    issn = {03770273}
}

@article{Hess1996,
    title = {{Viscosities of hydrous leucogranitic melts: A non-Arrhenian model}},
    year = {1996},
    journal = {American Mineralogist},
    author = {Hess, K-U and Dingwell, D D},
    pages = {1297--1300},
    volume = {81},
    url = {http://pubs.geoscienceworld.org/msa/ammin/article-pdf/81/9-10/1297/4326255/am81_1297.pdf},
    issn = {1945-3027}
}

@article{Giordano2008,
    title = {{Viscosity of magmatic liquids: A model}},
    year = {2008},
    journal = {Earth and Planetary Science Letters},
    author = {Giordano, Daniele and Russell, James K. and Dingwell, Donald B.},
    number = {1-4},
    month = {7},
    pages = {123--134},
    volume = {271},
    doi = {10.1016/j.epsl.2008.03.038},
    issn = {0012821X}
}

@article{Dingwell1996,
    title = {{Volcanic Dilemma: Flow or Blow?}},
    year = {1996},
    journal = {Science},
    author = {Dingwell, D. B.},
    month = {8},
    pages = {1054--1055},
    volume = {273},
    url = {https://www.science.org}
}

@article{Plank2013,
    title = {{Why do mafic arc magmas contain {\~{}}4wt{\%} water on average?}},
    year = {2013},
    journal = {Earth and Planetary Science Letters},
    author = {Plank, Terry and Kelley, Katherine A. and Zimmer, Mindy M. and Hauri, Erik H. and Wallace, Paul J.},
    month = {2},
    pages = {168--179},
    volume = {364},
    doi = {10.1016/j.epsl.2012.11.044},
    issn = {0012821X}
}

\appendix
\section{Non-dimensionalization}
\label{appendix:nondimensionalization}
We non-dimensionalize the equations using the standard choice for viscosity-dominated pressure where: 
\begin{subequations}
    \begin{align}
        r &= Lr^\prime \: , \\
        z &= Lz^\prime \: ,  \\
        u &= Uu^\prime \: , \\
        P &= \frac{\bar{\mu}U}{L}P^\prime \: , \\
        P_0 &= \frac{\bar{\mu}U}{L}P_0^\prime \: , \\
        t &= \frac{L}{U} t^\prime \: , \\
        \rho &= \bar{\rho}\rho^\prime \: , \\
        \eta &= \bar{\mu}\eta^\prime \: , \\
        \beta &= \frac{L}{\bar{\mu}U}\beta^\prime \: , \\
        \phi &= \phi^\prime \: , \\
        g &= \bar{g}g^\prime \: ,
    \end{align}
\end{subequations}
for a characteristic length scale, $L$, which is the maximum range of the spatial coordinate in both geometries; a characteristic velocity scale, $U$, which we choose to be the L$_2$-norm of the velocity field in the preceding time step; a characteristic viscosity, $\bar{\mu}$, which we choose to be the mean melt viscosity; a characteristic density, $\bar{\rho}$, which we choose to be the mean melt density; and for the cylindrical geometry, a characteristic gravitational acceleration, $\bar{g}$. \par 
We arrive at the well-known non-dimensional numbers: 
\begin{subequations}
    \begin{align}
        \mathrm{Re} &= \frac{\bar{\rho}UL}{\bar{\eta}} , \\
        \mathrm{Fr} &= \frac{U}{\sqrt{\bar{g}L}}
    \end{align}
\end{subequations}

We substitute into the equations for mass and momentum balance for the spherical case: 
\begin{subequations}
\begin{align}
    \frac{\partial \rho^\prime}{\partial t^\prime} &+ \frac{1}{r^{\prime 2}} \frac{\partial}{\partial r^\prime} \left( \rho^\prime r^{\prime 2} u^\prime \right) = 0 \: , \\
   \mathrm{Re} \frac{\partial u^\prime}{\partial t^\prime} &= - \frac{1}{\rho^\prime}\frac{\partial P^\prime}{\partial r^\prime} + \frac{4}{3} \frac{\eta^\prime}{\rho^\prime} \left( \frac{1}{r^{\prime 2}} \frac{\partial}{\partial r^\prime} \left( r^{\prime 2} \frac{\partial u^\prime}{\partial r^\prime} \right) - \frac{2 u^\prime}{r^{\prime 2}} \right) \: , 
\end{align}
\end{subequations}

and cylindrical confined geometry: 
\begin{subequations}
\begin{align}
    \frac{\partial \rho^\prime}{\partial t^\prime} &+ \frac{\partial}{\partial z^\prime} \left( \rho^\prime u^\prime \right) = 0 \: , \\
    \mathrm{Re} \frac{\partial u^\prime}{\partial t^\prime} &= - \frac{1}{\rho^\prime}\frac{\partial P^\prime}{\partial z^\prime} + \frac{\eta^\prime}{\rho^\prime} \left( \frac{4}{3} \frac{\partial^2 u^\prime}{\partial z^{\prime 2}} - 4 \frac{L^2}{R^2} u^\prime \right) + \frac{\mathrm{Re}}{\mathrm{Fr}^2} g^\prime \: , 
\end{align}
\end{subequations}

\cleardoublepage 



\end{document}